\documentclass[12pt,a4paper]{article}   	% use "amsart" instead of "article" for AMSLaTeX format

\usepackage{amsmath,amssymb,amscd}
\usepackage{graphicx}% Include figure files
\usepackage{dcolumn}% Align table columns on decimal point
\usepackage{bm}% bold math

\usepackage[left=2cm,right=2cm,top=2cm,bottom=2cm]{geometry}

\usepackage[colorlinks,allcolors=blue]{hyperref}

\providecommand{\href}[2]{\texttt{#2}}
\providecommand{\url}[1]{\texttt{#1}}

\newcommand{\half}{\frac{1}{2}}
\newcommand{\bea}{\begin{eqnarray}}
\newcommand{\eea}{\end{eqnarray}}
\newcommand{\beq}{\begin{equation}}
\newcommand{\eeq}{\end{equation}}
\newcommand{\nnel}{\nonumber \\ {}}

\providecommand{\eqref}[1]{equation (\ref{#1})}
\newcommand{\figref}[1]{Fig.\ \ref{#1}}

\newcommand{\secref}[1]{Section \ref{#1}}
\newcommand{\appref}[1]{Appendix \ref{#1}}
\newcommand{\tabref}[1]{Table \ref{#1}}

\newcommand{\tnote}[1]{} %{\textbf{(T)}\footnote{\textbf{(T)} #1}}
\newcommand{\tcomment}[1]{} %{\textbf{(T)} #1 \textbf{(end of T)}}
\newcommand{\vnote}[1]{} %{\textbf{(V)}\footnote{\textbf{(V)}#1}}
\newcommand{\vcomment}[1]{} %{\textbf{(V)} #1 \textbf{(end of V)}}

\newcommand{\tpre}[1]{}
\newcommand{\tprenote}[1]{}

\providecommand{\href}[2]{{#2}\footnote{See \texttt{#1}}}
\newcommand{\tseurl}[1]{\texttt{\url{#1}}}

\newcommand{\unitmat}{\hbox{\textsf{1}\kern-.25em{\textsf{I}}}}

\newcommand{\kin}{k^{(\mathrm{in})}}
\newcommand{\kout}{k^{(\mathrm{out})}}

\newcommand{\kouteff}{k^{(\mathrm{out})}_\mathrm{eff}}

\newcommand{\mrnd}{m_\mathrm{rnd}}

\newcommand{\pbar}{\bar{p}}

\newcommand{\tbar}{\bar{t}}

\newcommand{\Ecal}{\mathcal{E}}
\newcommand{\Gcal}{\mathcal{G}}

\newcommand{\Ncal}{\mathcal{N}}

\newcommand{\Npast}{\Ncal^{(-)}}

\newcommand{\Pcal}{\mathcal{P}}
\newcommand{\Pcalh}{\Pcal_{\mathrm{height}}}

\newcommand{\Pcalhat}{\widehat{\Pcal}}
\newcommand{\Ppast}{P^{(-)}}

\newcommand{\Vcal}{\mathcal{V}}

\newcommand{\ra}{\rightarrow}
\newcommand{\Ra}{\Rightarrow}

\newcommand{\Zbb}{\mathbb{Z}}

\newcommand{\labelmax}{\mathrm{max}}
\newcommand{\labelgr}{\mathrm{gr}}
\newcommand{\labelobs}{\mathrm{obs}}

\newcommand{\agr}{a_{\labelgr}}
\newcommand{\amax}{a_{\labelmax}}
\newcommand{\aobs}{a_{\labelobs}}
\newcommand{\mugr}{\mu_\labelgr}

\newcommand{\Pimax}{\Pi_{\labelmax}}
\newcommand{\ellmax}{L} %{\ell^{\labelmax} }}
\newcommand{\ellgr}{\ell} %{\ell^{\labelgr} }}
\newcommand{\ellgrbar}{\bar{\ellgr}} %{\bar{\ell}^{\labelgr} }
\newcommand{\ellgrhat}{\hat{\ellgr}} %{\bar{\ell}^{\labelgr} }
\newcommand{\ellgrobs}{{\ellgr}^{\labelobs}}
\newcommand{\ellmaxobs}{{\ellmax}^{\labelobs}}

\newcounter{pointcounter}

\newcommand{\tdef}[1]{\textsc{#1}}

\newcommand{\texpect}[1]{\langle #1 \rangle}

\begin{document}

%%%\flushbottom
%%%\maketitle
%%%\thispagestyle{empty}

\begin{center}
{\Large\textbf{The Longest Path in the Price Model\footnote{This is a post-peer-review, pre-copyedit version of an article published in Scientific Reports. The final authenticated version is available online at: http://dx.doi.org/[insert DOI]". }
}} \\[\baselineskip]
 {\large \href{http://www.imperial.ac.uk/people/t.evans}{Tim S.\ Evans},
  {Lucille Calmon},
  {Vaiva Vasiliauskaite}
 }
 \\[0.5\baselineskip]
 \href{http://complexity.org.uk/}{Centre for Complexity Science}, and \href{http://www3.imperial.ac.uk/theoreticalphysics}{Theoretical Physics Group},
Imperial College London, SW7 2AZ, U.K.
\\[0.5\baselineskip]
 2nd March 2020
\end{center}
\begin{abstract}
The Price model, the directed version of the Barab\'{a}si-Albert model, produces a growing directed acyclic graph.
We look at variants of the model in which directed edges are added to the new vertex in one of two ways: using cumulative advantage (preferential attachment) choosing vertices in proportion to their degree, or with random attachment in which vertices are chosen uniformly at random.
In such networks, the longest path is well defined and in some cases is known to be a better approximation to geodesics than the shortest path.
We define a reverse greedy path and show both analytically and numerically that this scales with the logarithm of the size of the network with a coefficient given by the number of edges added using random attachment. This is a lower bound on the length of the longest path to any given vertex and we show numerically that the longest path also scales with the logarithm of the size of the network but with a larger coefficient that has some weak dependence on the parameters of the model.
\end{abstract}

\section*{Introduction}

The Price model \cite{P65a,P76} is one of the oldest network models and it was motivated by the pattern of citations in academic papers. In a citation network, each node represents a document while every entry in the bibliography of a document $t$ is represented by a directed edge from an older document, node $s$, to node $t$.    One of the key features of a citation network, one inherent in the Price model, is that there is a fundamental \emph{arrow of time} in the network; bibliographies can only refer to older documents. \tnote{and so our edges point forward in time, given our convention for their direction.}
This means that there are no cycles in the network, you can never find a path from a node that returns  to that node. Thus a citation network is an example of a Directed Acyclic Graph (DAG).

\begin{figure}[htb!]
\centering
\includegraphics[width=0.2\textwidth]{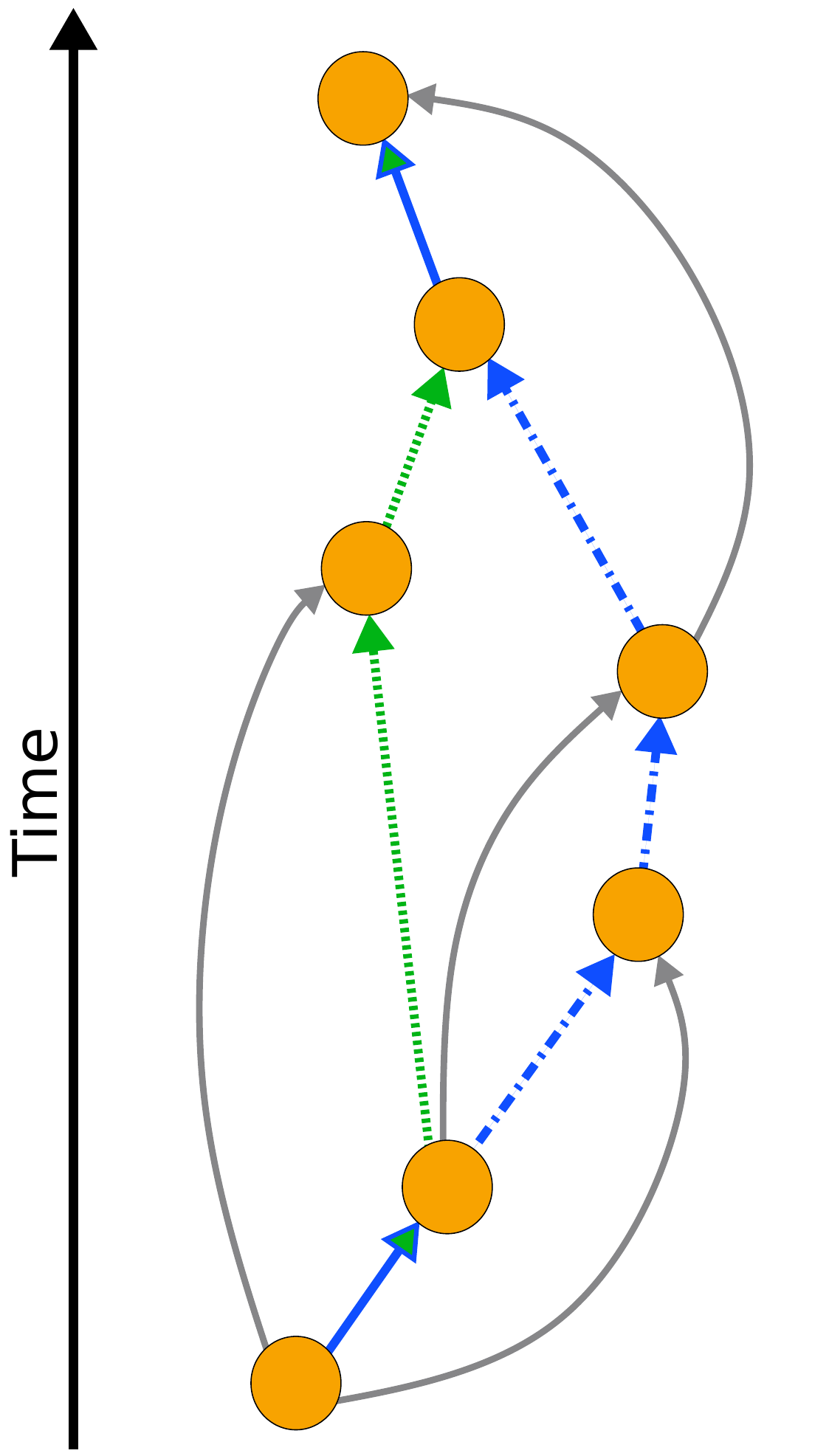}
\caption{An illustration of a Price-model style DAG where the longest, shortest and reverse greedy paths from last point to the first are distinct. The longest path from the source node to the sink node is highlighted in blue dot-dash line; the reverse greedy path is the dotted green path. Note the first edge is the same for both --- the green-blue edge. As illustrated here, the Price model produces DAGs which are neither transitively complete nor transitively reduced. In a transitively complete DAG, all nodes which are connected by a path are connected by a direct edge. Likewise, except for the case of one-incoming edge per node, the model is not transitively reduced \cite{CGLE14}, that is some edges could be removed without removing a path between any pair of nodes.}
\label{fpaths}
\end{figure}

Mathematically, DAGs have some distinctive properties and one of them is that for any pair of connected nodes there is a well defined and meaningful longest path length, for example see \figref{fpaths}.  Contrast this with, for example, undirected networks, where you can often find many paths between two given vertices that visit most of the nodes in a component so longest paths are often as long as the component is big, if all nodes in the path must be distinct, and infinite, if multiple visits to the same node were allowed. In directed graphs with cycles, the longest path is infinite, if multiple visits to a node are allowed. Both of these definitions of the longest path coincide if the network is acyclic, as the absence of cycles ensures that in any path, a node can only occur once. %As will be discussed in the following sections, the properties of longest path in a DAG

%%%fig 1

%The converse is that for most networks the shortest path is a useful concept, the six degrees of separation and so forth \cite{W03,N09b}. However i
In a citation network, it is not clear how useful the shortest path is.  For instance, in writing this paper, the oldest citation we have is to a paper by Price \cite{P65a}. The shortest path to Price's paper from this work has length one. On the other hand, most of the knowledge of that work contained in this paper did not come directly from that paper. We only reread Price's paper to check one detail while working on this project. So the length of the shortest path to that paper seems largely irrelevant. Rather, the information in this early bibliometrics paper by Price has reached us through a sequence of other work, much of it not explicitly referenced in our paper.  We drew on much more recent documents such as the reference book by Newman \cite{N09b} which in turn cites papers which developed various aspects of the Price model. Indeed there is much evidence \cite{SR05a,CGLE14,GAE14} that typically 70\% or so of a bibliography may not have been used directly when producing the work in an academic paper.

So our thesis is that for DAGs the longest path plays a much more important role than the shortest path.  In simple models the longest path has been shown to be the best approximation to the geodesic for models of DAGs embedded in Minkowski space \cite{BG91} where there is a single time direction\footnote{This has been exploited in real data sets where dimension and curvature of a DAG can be measured \cite{CE14,C17} enabling us to embed DAGs such as citation networks in Minkowski space \cite{CE16}.}.  A similar rigorous link for undirected networks has only been made for the shortest path in networks embedded in Euclidean space where there is no arrow of time \cite{EMY07,DMPP16}.
\tnote{Linking longest path to geodesics has been done in network models which have an ``arrow of time'' such as models embedded in Minkowski space and other non-Riemannian spaces. By way of contrast, the shortest path is linked rigorously  to geodesics in network models embedded in Riemannian spaces such as Euclidean space \cite{EMY07,DMPP16}.}

The properties of the longest path have been investigated in the context of simple models known as Cube Spaces \cite{BB91} which include those built from Poisson Point Processes in Minkowski spaces where all causally connected points are connected to form a network. However these are examples of transitively complete DAGs, that is if there is a path between two points then there is always an edge connecting those two points directly.  However, that is not true for a citation network where the limited size of a bibliography means no document ever cites every older paper to which it has some connection. What we seek to do in this paper is to look at the properties of the longest path in a simple model, the Price model \cite{P76} and its variations, where the network is neither transitively complete nor (except for one parameter value) transitively reduced, the situation for most DAGs in a social context such as citation networks.  Can we calculate the length of the longest path in the Price model? How does this length depend on the parameters of the model?

We start by outlining our analytic results.  We then compare these predictions to numerical simulations and summarise our findings.

% *********************************************************************
\section*{Analytic Results}\label{sanalyticsol}

%\begin{itemize}
%\item Define the Price model and the variation used here.
%\item Simple approach
%\item Quote full result
%\end{itemize}

\begin{figure*}[htb!]
\centering
\includegraphics[width=0.75\textwidth]{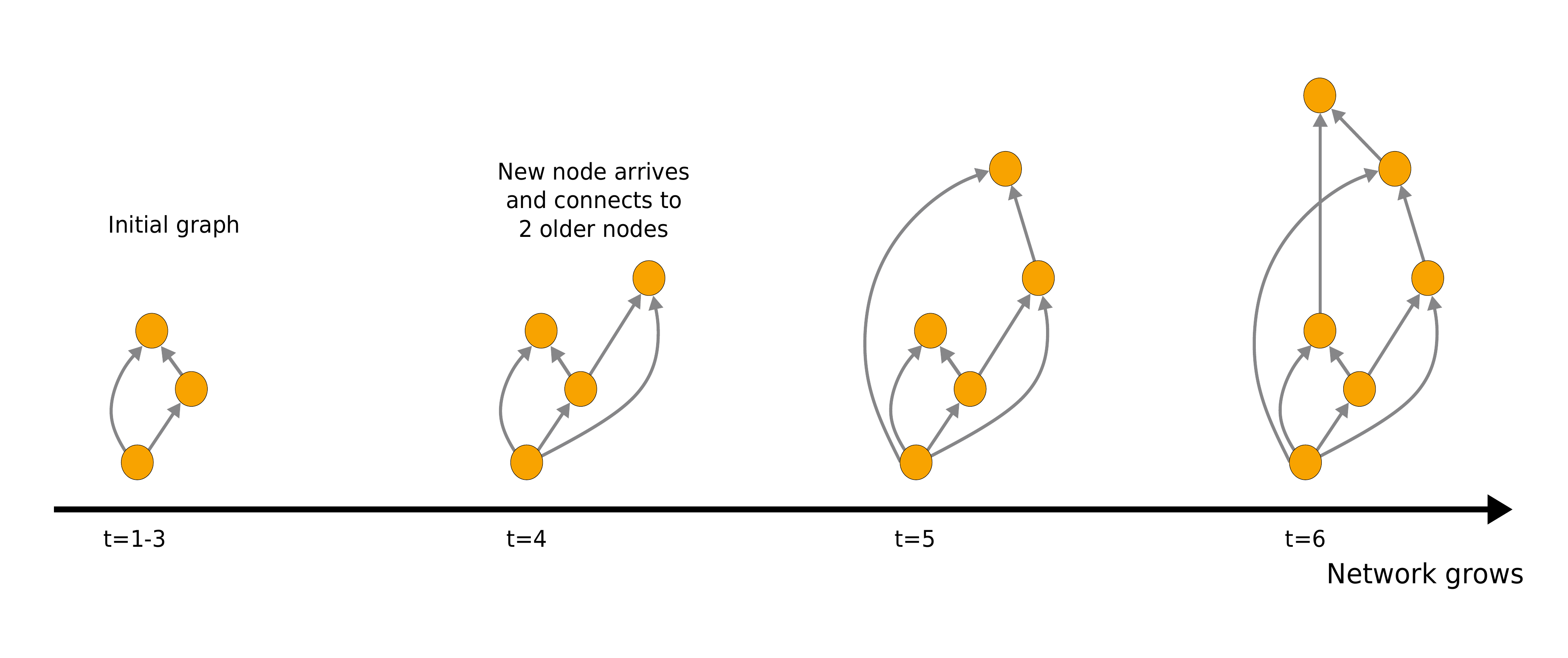}
% \tcomment{e.g.\ arrow reversed version of Fig.3 of \cite{C19}.}
\caption{An illustration of the Price Model. Here the height of the node on the page indicates the time with the first node at the bottom being the node $t=1$. At each stage we show the graph $\Gcal(t)$ so after the new nodes and its incoming edges have been added.
}
\label{fPrice}
\end{figure*}

In the Price model \cite{P76} (for instance see sec.14.1 of Newman \cite{N09b}) we start from a network $G(t)$ defined at an integer `time' $t$. We create a new graph $G(t+1)$ by first adding one new vertex, which we label with the time $(t+1)$.
%The initial time $t_0=1$ is the creation time and index of the first node.
This new node, $(t+1)$, is connected to $m$ existing vertices $s$ in the graph $G(t)$.  These $m$ existing vertices $\{s\}$ are each chosen with probability $\Pi(t,s)$. We will use a convention that these edges point from older to newer vertices, from $s$ to $(t+1)$. Once these edges have been added we have our new graph $G(t+1)$.  The process is then repeated. For an example of how a network grows according to the Price model, see \figref{fPrice}.
%It makes the first node node $t=t_0=1$ the only sink node.

The mathematical and numerical simplicity of this model comes from the simple definition of $\Pi(t,s)$. To define the probability $\Pi(t,s)$ we first define $N(t)=N_0 + t$ be the number of nodes in the graph $G(t)$ for some constant $N_0$. The number of edges in the graph $G(t)$, after all $m$ edges have been added to node $t$, is $E(t)=E_0 + mt$ where $E_0$ is some constant.
Finally in the graph $G(t)$ let the node created at time $s$ have out-degree $\kout(t,s)$, the number of edges leaving $s$ and connecting it to later nodes.

In this model,  the connection of edges to new node $(t+1)$ is made in one of two ways. With probability $p$ node $(t+1)$ is connected to an existing vertex $s$ chosen in proportion to the number of edges leaving $s$ for later nodes in $G(t)$,  $\kout(t,s)$.
Price called this \tdef{cumulative advantage} and, after normalisation, we have that the probability of choosing $s$ is $\kout(t,s)/E(t)$.  The second process happens with probability $\pbar=(1-p)$ and in this case we choose the source vertex $s$ uniformly at random from the set of vertices in $G(t)$, i.e.\ with probability $1/N(t)$.  If we start the process at time equal to $1$, then the probability of connecting the vertex $(t+1)$ to existing vertex $s$ is $\Pi(t,s)$ where
%\beq
%  \Pi(t,s) =
%  \left\{
%  \begin{array}{r l}
%  p \frac{\kout(t,s)}{E(t)} + \pbar \frac{1}{N(t)}
%  & \mbox{ if } t \geq s \geq 1  \; \mbox{ unless } \; t=s=1\, ,
%  \\
%  1 &  \mbox{ if } t=s=1 \, ,
%  \\
%   0 & \mbox{ otherwise}
%  \, .
%  \end{array}
%  \right.
%  \label{PricePidef}
\beq
  \Pi(t,s) =
  p \frac{\kout(t,s)}{E(t)} + \pbar \frac{1}{N(t)}
  \mbox{ if } t \geq s \geq 1  \, ,
  \label{PricePidef}
\eeq
unless $s=t=1$ when $\Pi(t,s)=1$, otherwise $\Pi(t,s)=0$.
Note that in his original paper, Price considered $p=m/(m+1)$ where $\Pi \propto \kout + 1$. This more general form for the attachment probability $\Pi(s,t)$ in \eqref{PricePidef} has been used in many related contexts since Price, see Newman \cite{N09b} for a review.
%%%fig 2

There is an issue about the starting point for this process.
The usual form for $\Pi$, the $t\geq s > 1$ form in \eqref{PricePidef}, leaves us with a problem for $\Pi(t=1,s=1)$ when looking at the attachment to the second vertex, $t=2$. Our solution is to demand that $\Pi(t=1,s=1)=1$. This fixes the cumulative probability $\Pi_\leq$ to have a consistent value which is in fact all we need for this calculation. However, we will also assume that the number of nodes and number of edges are given by $N(t)=t$ and $E(t)=mt$ respectively. This is only needed for $t\geq 2$ so in principle we must allow multiple edges between nodes starting with $m$ edges added between node $2$ and node $1$. Again this cannot be true for at least the first node at $t=1$.

%\tnote{An alternative solution is to start the calculation from $t_0>1$ and probably for each of calculation with $E(t)=mt$ and $N(t)=t$ holding for $t \geq 1$. We would only need to allow for a non-zero $\ell(t=1)=\ell_0$ and this would give some control over initial conditions.}

We also note that our analytic calculations allow our networks to contain multiple edges (node pairs linked by more than one edge).  Of course, a real citation network and many numerical calculations of this model (though not our numerical calculations) do not have multiple edges. However, in the long time limit the effect of such edges becomes negligible as they form a small fraction of the edge population, a fraction that dies off as a power law in time \cite{N09b}.

Now we would like to define the longest path algebraically.  Unfortunately, finding the longest path requires global knowledge of all the paths.  This is extremely hard to do algebraically (though is surprisingly straightforward numerically). So the first stage of our calculation is to decide to calculate a path defined with local knowledge only.  That is we define what is called a \tdef{reverse greedy path} using an iterative process where at each stage we only need to know about the properties of the next vertex in the path.\tnote{A similar simplification is needed in the cube space calculations.} We will denote the length of the reverse greedy path from the source vertex $s=1$ to a target vertex $t$ as $\ellgr(t)$. The length of the longest path from the source vertex to a target vertex $t$ will be denoted as $L(t)$.

The reverse greedy path to a node $t$ is a path running from the source node at the initial time $t=1$ to node $t$. This always exists and it is unique.
To define it suppose that we have found the reverse greedy path to all earlier nodes.  The last step on the reverse greedy path to node $t$ is made along the edge arriving at $t$ from its most recent predecessor node, say $s$. The idea is that the most recent predecessor of node $t$, furthest from the source node in terms of the time,  is also the most likely to be the predecessor node furthest from the sink node in terms of network path lengths. There is no guarantee that our reverse greedy path is identical to the longest path, so the reverse greedy path length is a lower bound on the longest path length. A more formal definition is given in Appendix A.1 in the Supplementary Information. %%%\appref{agreedydef}.

Of course in any one instance of the Price model, this reverse greedy path length will fluctuate if we look at nodes of similar ages, not least because $\ellgr$ is integer valued. We will use a mean field approach so our $\ellgr(t)$ is an average over many realisations of the model though for simplicity we will not include the expectation value notation $\texpect{\ldots}$. For that reason our $\ellgr(t)$ will be a real valued  monotonically increasing function of time $t$.

We can find the long-time behaviour using the following simple argument. On average, there are $p m$ edges added with cumulative advantage at each time step.  Suppose we are adding a new node at time $(t+1)$ and we are looking for source nodes $s$ for the $m$ edges we are going to add. The source nodes chosen with cumulative advantage are those with largest degree and those are the oldest nodes created at small values of time $s$ (since $\kout(t,s) \propto (t-s)^p$, for instance see Newman \cite{N09b}). So the youngest source node chosen, nodes created at the largest value of time $s$, is likely to be one of the $\mrnd = m \pbar$ nodes chosen uniformly at random.
The probability that all these $\mrnd$ randomly chosen source nodes are chosen between time $1$ and time $\hat{s}$ inclusive is $(\hat{s}/t)^{\mrnd}$. Suppose we consider the time $\hat{s}_{1/2}$ where with probability one half the time coordinate of the largest randomly chosen source node is $\hat{s}_{1/2}$ or less, then this sets the scale for the birth date of the youngest source node connected to node $(t=1)$, namely that $\hat{s}_{1/2} = \mu t$ where $\mu=2^{-\mrnd}$. This is the previous node on the reverse greedy path from the initial node to node $t$. We can then estimate the numbers of steps it takes to get back to the source node at $t=1$ as
$\mu^{\ellgr} t \approx 1$ which leads to
\beq
 \ellgr(t) = \frac{m \pbar}{\ln(2)} \ln( t ) \, .
 \label{ellasoleasy}
\eeq

The simplicity of the attachment probability in the Price model means we can also produce more detailed derivation using a mean-field approach. Let the probability that the length of the reverse greedy path, $\ell$, from new node $(t+1)$ to the initial node at $t=1$, be $P(\ell,t)$. Then the master equation is of the form
\bea
 P(\ell,t+1)
 &=&
 \sum_{s=1}^{t} P(\ell-1,s) \Pimax(t,s) \, .
 \label{eGreedyMaster}
\eea
Here $\Pimax(t,s)$ is the probability that of the $m$ predecessor nodes connected to a new node at $(t+1)$, the oldest of them is $s$. In terms of the generating function $G(z,t) = \sum_{\ell=0}^\infty z^\ell P(\ell,t)$ we find that
the exact solution in the Price model is (see %%%\appref{aspricemaster}
Appendix A.3 in the Supplementary Information for details)
\bea
 G(z,t)
 &=&
 \prod_{s=1}^{t-1} \left( z \left( 1  - \left( \frac{s-\pbar}{s}  \right)^m\right)
  +
  \left( \frac{(s-\pbar)}{s} \right)^m
  \right) .
  \label{eGsol}
\eea
Exact forms for the expected reverse greedy path length can be found from this expression, especially for specific small values of $m$.  However, the leading order contribution for large times is always of the form
\bea
\lim_{t \ra \infty} \ell(t) &=&
   m\pbar \ln(t) - m\pbar \psi(m\pbar + 1)
  %%%\nonumber \\ &&
  + \sum_{n=2}^{m} \binom{m}{n}
  (-1)^{n-1} (\pbar)^n \zeta(n)
 + O(t^{-1})
 \label{eellasymp2}
 %- \frac{m\pbar (1/2)-m\pbar}{t}   + O(t^{-2} \label{aellasymp2}
\eea
where it is implicit that there is no contribution from the term with the sum for the case of $m=1$. Here $\psi(z)$ is the digamma function and $\zeta(z)$ is the Riemann zeta-function.
The details of the calculation are given in Appendix A.3 of the Supplementary Information.%\appref{aspricemaster}.

Finally, the scaling properties of the longest path in the Price model suggests that the properties of \tdef{height antichains} are also very simple. The \tdef{height} of a node in a DAG is the length of the longest path to a node from a source node, any node with zero in-degree.  Thus in the Price model, the height of a node is simply the length of the longest path length from the initial node to the given node, our $\ellmax$.

Nodes connected by a path cannot be of the same height.  Thus the subset of all nodes at the same height form an \tdef{antichain}, a set of nodes in which no two are connected by a path  \cite{VE20}. The scaling properties of these height antichains are simple to estimate if we conjecture that the average longest path $\ellmax$ of a node $t$, its average height, scales as $\ln(t)$. This suggests that if the median index of a node in an antichain of integer valued height $h$ is $t_\mathrm{mid} = (\mu)^h$ then the mean index of nodes in the antichain will scale as $\cosh(\sqrt{\mu})(\mu)^h$, the variance in the index of nodes in the antichain will be roughly $(1/\sqrt{3}) \sinh(\sqrt{\mu}) (\mu)^h$, and the number of nodes in the antichain will vary as in the $2\sinh(\sqrt{\mu}) (\mu)^h$.

% *********************************************************************
\section*{Numerical Methods and Results}

%\begin{itemize}
%\item Numerical simplifications and issues: multiedges, initial graph, algorithm used.
%\item reverse greedy path  from $s=1$ to $t$ length $\ellgr(t)$.
%\item Longest Path from $s=1$ to $t$ length $\ellmax(t)$
%\item $m$ dependence: $m=1$ (exact analytic results), $m>1$.
%\item $p$ dependence, including $p=0$ RA, $p=1/(1+m)$ (BA limit)
%\item dependence on initial graph, one example?
%\end{itemize}

In the master equation \eqref{eGreedyMaster}, multiedges (attaching two edges from the new vertex to the the same vertex) are not excluded.  In our numerical implementation code we also allowed multiedges to be created. However the probability of attaching one edge from new vertex $(t+1)$ to any existing vertex $s$ is decreasing as $(s/t)^{p}$ (for instance see p.\ 489-90 Newman \cite{N09b}). So the creation of a multiedge becomes negligible at large times hence our networks are essentially the same as implementations of the Price model in which multiedges are excluded.

The first few steps of the numerical implementation of the Price model have some subtleties which are worth mentioning.
%The Price model is awkward to implement numerically for the first few steps.
The problems noted analytically with the initial node at $t=1$, which is the only node with zero-in-degree, exemplify the issue.
%If multiedges are allowed, then adding a second node with $m$ incoming edges sets the out-degree of node 1 to be $m$ at $t=2$ and makes the degree distribution very skewed.
%We have assumed $E(t)=mt$ but that is incompatible with the in-degree of each node (except the first at $t=1$) being $m$.  Of course, in the long-time limit these issues become less important as almost all nodes have in-degree $m$ and so the number of edges scales as $mt$ asymptotically. For similar reasons, degree distribution follows the long-time limit predicted from mean-field master equations extremely well, except for the finite size effects seen in the cutoff of the distribution at large degree.  The initial graph has a major influence on the degree of the very
The earliest nodes are those with the shortest values of our $\ellgr$ and $\ellmax$ path lengths to the first node. Since the path lengths of the first few nodes will be added to any other path routed via one of these early nodes, we expect the initial graph to give a constant contribution to the path lengths we measure but not to alter the growth in length scales over long-times.

We chose to start our simulations from a complete graph of $(2m+1)$ nodes, labelled $t=1$ to $t=(2m+1)$. All pairs of nodes are connected in this initial graph, with the edge direction from earlier to later node.
This initial graph ensures that $E(t)=mt$ for all graphs generated numerically,  $G(t)$ for $t\geq (2m+1)$. The out-degree distribution is fairly even but the in-degree is not fixed to be $m$ for nodes in this initial graph.
%The reverse greedy path and longest path lengths for each node in this initial graph are easily calculated at the start of the numerical simulation.
Further notes on the effect of the initial graph are given in Appendix B.4 in the Supplementary Information.
%%% \appref{ainit}.

%For large times, the chance of attaching a new node to a node the initial graph via either cumulative advantage or random attachment. So we expect to be able to neglect the effect of the initial graph for large times.

In order to simplify and accelerate the numerical analysis, for each new node $(t+1)$ added we drew nodes uniformly at random from an ``attachment list'' which we maintain. After we have chosen the source edges for the $m$ edges attached to the new vertex, we update our attachment list by adding each source node once for every edge, and we add $(\pbar/pm)$ references to the new node. This means we restrict our results to cases where $(\pbar/pm)$ is an integer.
%The length of the list is $m(1+p/\pbar)t$ after node at time $t$ has been added indicating there is also a cost in terms of memory for our approach. However it ensures that
Drawing nodes uniformly from our attachment list means that we are choosing vertices according the the probability \eqref{PricePidef}. For the special case where there was no cumulative advantage, $p=0$, the attachment list was simply a list of the existing vertices where each is referenced once. A more detailed explanation is given in Appendix B.1 of the Supplementary Information.
%%% \appref{apricealg}.

For each node $s$, we also record values for the lengths of the reverse greedy path $\ellgr(s)$ and the longest path $\ellmax(s)$.  When adding a new node $t$, it is simple to look at the values of the lengths of these paths to the $m$ nodes attached to the new node. From that information, it is simple to record the lengths of these paths to the new vertex, $\ellgr(t)$ and $\ellmax(t)$. Storing and manipulating these results proved to be more of a limiting factor than the speed to produce them. We produced results for networks of up to $10^8$ nodes.
%The limiting factor was not the speed to produce the networks but rather the size of the data which meant storage, transfer and analysis of results was the bottleneck.

%%%fig 3
\begin{figure*}[htb!]
    \centering
    \begin{tabular}{p{0.475\textwidth}p{0.475\textwidth}}
        %\verb|m_eq_2_a_eq_1_N_eq_100000000_step_no_1_lengths_greedy_longest_path_fit.png|
        %\includegraphics[width=0.475\textwidth]{m_eq_2_a_eq_1_N_eq_100000000_step_no_1_lengths_greedy_longest_path_fit.png}
        \includegraphics[width=0.475\textwidth]{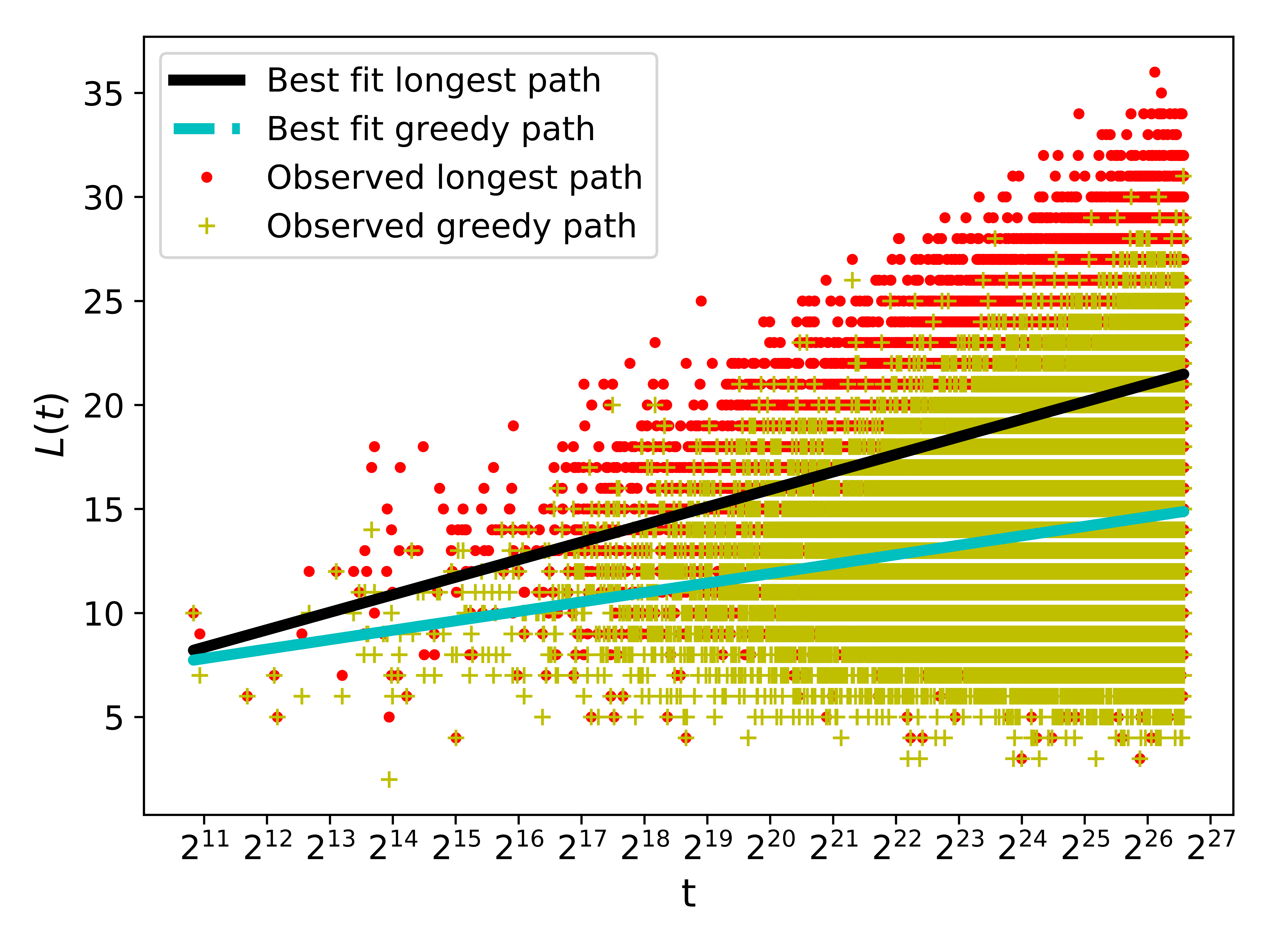}
        %\label{fonerun}
        &
        %\verb|m_eq_5_a_eq_3_N_eq_100000000_number_of_runs_eq_100_greedy_longest_path_fit_cutoff_eq_100000000_samplesize_eq_100000.png|
        \includegraphics[width=0.475\textwidth]{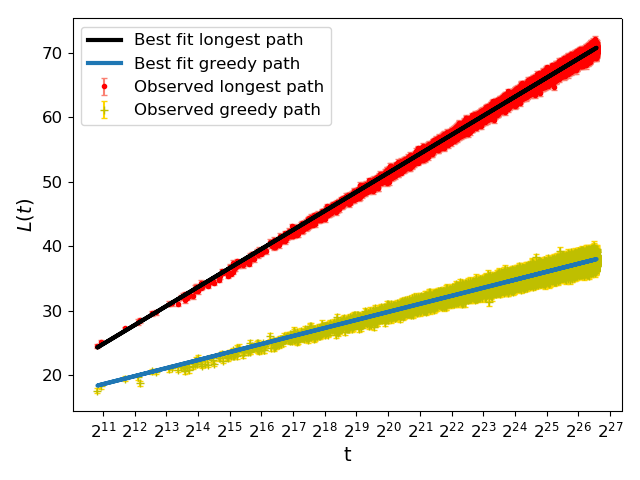}
        %\label{fmultirun}
        \\
        {{\small \textbf{(a)} $m=2$, $\alpha=m\pbar/p=1$, $N=10^8$, single run.}}
        &
        {{\small \textbf{(b)} $m=5$, $\alpha=m\pbar/p=3$, $N=10^8$, averaged over 100 runs (a random sample of $10^5$ points is plotted).}}
    \end{tabular}
        \caption{On the left is a plot of path length against $t$ for a single network realisation showing noisy the data is.  Averaging over 100 runs greatly reduces the fluctuations as shown in the righthand plot. Fitted lines are of the form $a \ln(t) +b$. In both figures, a random sample of $10^5$ points is plotted. The abscissa axes are logarithmic to show the linear behaviour between the path lengths and $t$.}
\label{frunfluct}
\end{figure*}

The results for these path lengths are quite noisy for any one node as shown by an exemplary run in \figref{frunfluct}.
%\figref{fonerun}.
Despite the relatively large fluctuations in results for any one node,there is a clear trend in the nodes created at later times.
%This is due to the fact that the path lengths are not independent in one network realisation.
The fluctuations of the path length scaling are greatly reduced when averaged over multiple networks as shown in \figref{frunfluct}. So we use 100 runs for each set of parameter values in our work.

%%%%%%%%%%%5
%Figure 3
%%%%%%%%%%%

%%%%\iffalse%%%
%\begin{figure*}[htb!]
%    \centering
%   % \includegraphics{longestpathpriceFig3}
%   \begin{tabular}{p{0.475\textwidth}p{0.475\textwidth}}
%        %\verb|m_eq_2_a_eq_1_N_eq_100000000_step_no_1_lengths_greedy_longest_path_fit.png|
%        \includegraphics[width=0.475\textwidth]{m_eq_2_a_eq_1_N_eq_100000000_step_no_1_lengths_greedy_longest_path_fit_sample_size_eq_100000.pdf}
%%        %\label{fonerun}
%        &
%        %\verb|m_eq_5_a_eq_3_N_eq_100000000_number_of_runs_eq_100_greedy_longest_path_fit_cutoff_eq_100000000_samplesize_eq_100000.png|
%        \includegraphics[width=0.475\textwidth]{m_eq_5_a_eq_3_N_eq_100000000_number_of_runs_eq_100_greedy_longest_path_fit_cutoff_eq_100000000_samplesize_eq_100000.png}
%%        %\label{fmultirun}
%        \\
%        {{\small \textbf{(a)} $m=2$, $\alpha=m\pbar/p=1$, $N=10^8$, single run.}}
%        &
%        {{\small \textbf{(b)} $m=5$, $\alpha=m\pbar/p=3$, $N=10^8$, averaged over 100 runs.}}
%    \end{tabular}
%        \caption{On the left is a plot of path length against $\ln(t)$ for a single network realisation showing noisy the data is. Averaging over 100 runs greatly reduces the fluctuations as shown in the righthand plot . Fitted lines are of the form $a \ln(t) +b$. In both figures, a random sample of $10^5$ points is plotted.}
%\label{frunfluct}
%\end{figure*}
%%%%\fi%%%

\begin{figure*}[htb!]
\centering
\includegraphics[width=\linewidth]{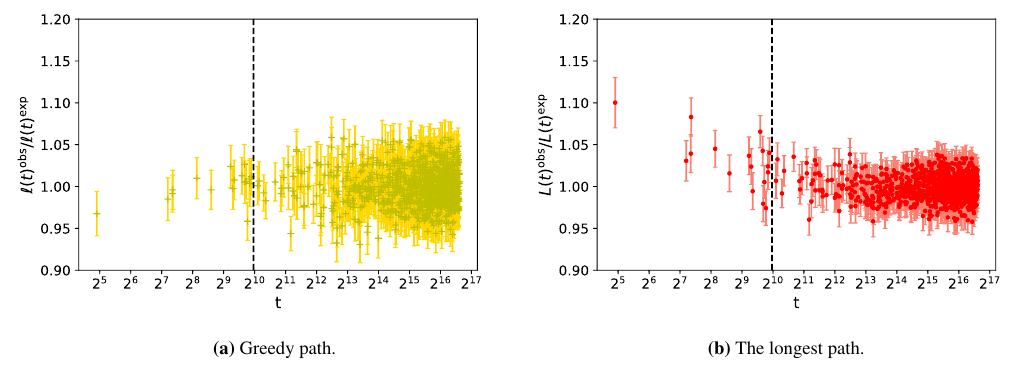}
%    %\begin{tabular}{p{0.48\textwidth}p{0.48\textwidth}}
%     \begin{tabular}{cc}
%        %\verb|m_eq_5_a_eq_3_N_eq_100000000_number_of_runs_eq_100_longest_path_obs_fit_ratio_cutoff_eq_100000000_samplesize_eq_1000000_loglin_ylimits.png|
%        \includegraphics[width=0.45\textwidth]{m_eq_5_a_eq_3_N_eq_100000000_number_of_runs_eq_100_longest_path_obs_fit_ratio_cutoff_eq_100000000_samplesize_eq_1000000_loglin_ylimits.png}
%        &
%        %\verb|m_eq_5_a_eq_3_N_eq_100000000_number_of_runs_eq_100_greedy_path_obs_fit_ratio_cutoff_eq_100000000_samplesize_eq_1000000_loglin_ylimits.png|
%        \includegraphics[width=0.45\textwidth]{m_eq_5_a_eq_3_N_eq_100000000_number_of_runs_eq_100_greedy_path_obs_fit_ratio_cutoff_eq_100000000_samplesize_eq_1000000_loglin_ylimits.png}
%        \\
%        {{\small \textbf{(a)} The longest path.}}
%        &
%        {{\small \textbf{(b)} Reverse greedy path.}}
%        \end{tabular}
\caption{The ratio of the average over 100 runs of the path length measured numerically, $\ellmaxobs$ and $\ellgrobs$, divided by the expected values, as described by the numerical best fit. The data for $p=0.375$ $m=5$ from $t=1000$ (represented by a vertical dashed line) to $t=10^8$ was fitted to \eqref{fitequation}. The reverse greedy path results are shown on the left, and the longest path are shown on the right. The error bar on each point is calculated from the standard error in the mean from the results for each node over 100 runs. A random sample of $10^3$ points is plotted in both figures.}
%Note that we have only shown the length for one node in every 100 nodes in order to show the fluctuations and trend clearly.}
%\tcomment{Version of Fig.11,12 of \cite{C19}.}
\label{foneparamfit}
\end{figure*}

In order to compare our numerical data with the analytical results we fitted the path lengths found to the function $f(t)$ where
\beq
 f(t) = a\ln(t) + b \, .
 \label{fitequation}
\eeq
The fit was made by using a non-linear fitting routine based on the optimisation of the chi-squared measure of goodness of fit (for instance see \cite{PTVF92}) as described in more detail in Appendix B.3 of the Supplementary Information. Errors on parameters were estimated from the covariance matrix  produced by such a method. 
Given that our analytical work only studies the long-time limit and that the early times in the numerical simulation do not satisfy all the conditions of the analytical work, it is not surprising that in \figref{foneparamfit} we still see significant difference between the numerical results and analytical predictions for the length of paths from $t=1$ to those nodes created at early times.  So when fitting to our numerical data we only use data for nodes created from time $t_0=1,000$ up to the last node at $t=10^8$. The effect of this cutoff is discussed in Appendix B.3 but we found varying this lower cutoff had little effect on our results since we had so many data points from the region where the asymptotic growth dominates.
%%%fig 4

\begin{figure}%[htb!]
\centering
%\verb|a_obs_divided_by_m_pbar_various_m_p.png|
\includegraphics[width=.6\linewidth]{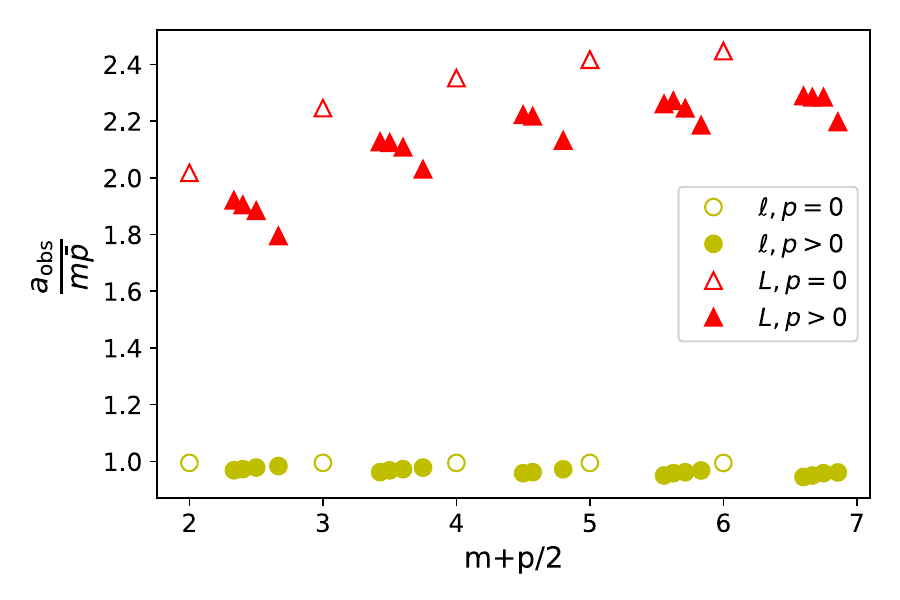}
\caption{The ratio of $\aobs/(m\pbar)$ where $\aobs$ is the coefficient of $\ln(t)$ derived from the best fit of the numerical path length data to $a\ln(t) + b$ \eqref{fitequation} while $m\pbar$ is the analytical prediction for the value of $a$ when looking at the length of the reverse greedy path.  The red triangles show the results for the reverse greedy path value of $a$ while yellow circles are the longest path values.
%The horizontal axis is $m+p/2$.
These values were obtained by fitting the form to nodes created between $t=1,000$ and $t=10^8$ from 100 realisations.
Errors on the fitted values of $a$ were smaller than the marker size and are not shown.
}
\label{fallparamafit}
\end{figure}

\begin{figure}%[htb!]
\centering
%\verb|a_max_divided_by_a_gr_various_m_p.png|
\includegraphics[width=.6\linewidth]{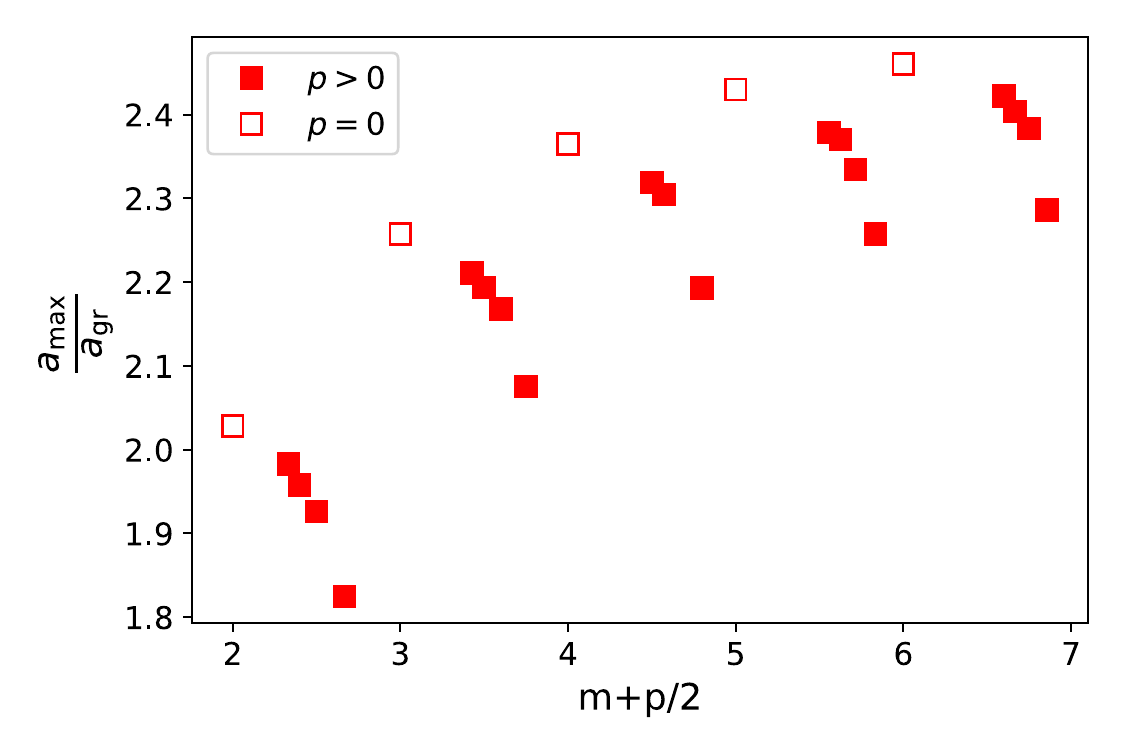}
\caption{The ratio of $\amax/\agr$ where $a$  is the coefficient of $\ln(t)$ in the best fit of the numerical path length data to \eqref{fitequation}, $\amax$ for the longest path data and ${\agr}$  for the reverse greedy path data.
%The horizontal axis is $m+p/2$.
These values were obtained by fitting the form to nodes created between $t=1,000$ and $t=10^8$ from 100 realisations. As a result the errors on the fitted values of $a$, as estimated from the covariance matrix of the linear fitting algorithm, were smaller than the marker size and so these are not shown.}
\label{fallparamaratiofit}
\end{figure}

The dependence of the coefficient of the $\ln(t)$ term found from the fit, $a$, on the model parameters is shown in \figref{fallparamafit} and \figref{fallparamaratiofit}.
%%%fig 5

%%%fig 6

The next-to-leading order coefficient, $b$ of \eqref{fitequation}, showed no clear trends.
We also considered a non-linear fit with a term of $c/t$ added to the expression in the \eqref{fitequation}. We found that in practice, this term had little influence on the remaining parameters of interest, namely, $a$ and $b$. Furthermore, the errors in $c$ were found to be relatively large in comparison to the errors of the parameters $a$.

Finally, it is clear that the longest path length is scaling as $\ln(t)$ to a good approximation. As noted above this then implies that the properties of height antichains in the Price model should follow a regular pattern which depends on the height of nodes in each antichain.  Numerical confirmation of these patterns are given in Appendix B.6 of the Supplementary Information.
%%% \appref{anumantichain}.

\clearpage

% *********************************************************************
\section*{Discussion}\label{sdiscussion}

%\begin{itemize}
%\item Motivation for $\Pi$, random walk papers
%\end{itemize}

The numerical results for the leading behaviour of the path length scales are striking.  Within the margin of numerical error, our results in \figref{fallparamafit} show that the length of the reverse greedy path scales asymptotically as $m\pbar \ln(t)$ for a wide range of parameter values.  This is consistent with both the simple argument and the detailed analytical calculation presented in the Analytic Results (see also Appendix A of the Supplementary Information). The analytical approach also shows that for long times, the distribution of lengths of reverse greedy paths in the Price model is Poisson distributed with mean equal to $m\pbar \ln(t)$ (see Appendix A).  %In particular, the variance of the path length is equal to the mean, but we have not investigated the variance numerically here.

The reverse greedy path length is a lower bound on the length of the longest path so it is no surprise that the longest path length also scales as $\ln(t)$ with a coefficient, $\amax$, which is larger than the corresponding scaling factor for the reverse greedy path length, $\agr$.  Interestingly this coefficient of the $\ln(t)$ term, $\amax$, for the longest path shows some additional weak dependence on the parameters beyond the $m\pbar$ found for the reverse greedy path, as both \figref{fallparamafit} and \figref{fallparamaratiofit} clearly show.

The Price model is not in itself a very realistic model for any particular context.  For instance, a true citation network often shows many other features such as a preference to cite recent papers, for example see \cite{GS94,GS95,R00,WFC14,GAE14,MWBRS16,GS17}.  The choice of a simple linear form for the attachment probability \eqref{PricePidef} appears to be part of this simplification, a form linear in degree motivated by the need for mathematical simplicity.
At first sight, this form seems unrealistic since it requires authors of papers to have global information about the citation network because of the normalisation factors.  No author can know exactly how many citations a paper has let alone the total number of citations in the network. However, this form emerges naturally in many situations as the result of doing local searches on the network, see \cite{V01a,KR01,SK04,ES05,JR07,SR07a,GAE14,GS17} and references therein.
In more realistic models, the cumulative advantage, the $p$ term in $\Pi$ emerges from doing a local search back through the current citation network, while the random attachment, the $\pbar$ term in $\Pi$, represents a simple model of other possible processes. So like all good models, the emphasis in the Price model on the linear form for $\Pi$ in terms of degree does capture an important and realistic feature of many real situations.
This linear form of the attachment probability $\Pi$ is also the critical feature in the analysis of  undirected versions of this model, such as the Barab\'{a}si-Albert model \cite{BA99} where the cumulative advantage aspect is known as preferential attachment and the original example worked with $p=1/2$ in our notation.

However, the Price model, simple as it is, also emphasises another critical aspect of a citation network, and that is the inherent arrow of time in this context. Citations (almost) always point backwards in time\footnote{Typical data sets \cite{CGLE14,CE14} suggest that less than 1\% of citations are to documents which are labelled as being published later than the citing document.} and the networks created in the Price model are realistic in this way, they always produce directed acyclic graphs.
This acyclic property is lost when the edge direction is ignored, as in the Barab\'{a}si-Albert implementations of this model. Since many analyses work in the undirected version, they have missed this key feature of the Price model and of real-world citation networks.

For instance, the length of the shortest path between two nodes is a natural measure for undirected networks since in some circumstances it can be related to the geodesic of networks embedded in Euclidean space, for example see \cite{EMY07,DMPP16}.
For an undirected version of the Price model, the LCD model of Bollob{\'a}s and Riordan \cite{BR04,B03c} (a more precisely defined version of the Barab\'{a}si-Albert model \cite{BA99}), it is known that the diameter, the largest length of any shortest path between two nodes, scales as $\ln(t)/(\ln\ln(t))$ if $m>1$ while the diameter scales as $\ln(t)$ for the special case of $m=1$ \cite{BR04} (see also theorem 18 of \cite{B03c}).
For the case $m>1$, Bollob{\'a}s points out that while in any random graph we expect to see the small-world effect \cite{WS98} and a $\ln(N)$ scaling of lengths (for $N$ nodes in the network), for the undirected version of this model ``one might expect the
diameter to be even smaller: the unbalanced degree distribution pushes up the number of small paths, and thus, perhaps, pushes the diameter down'' (see Bollob{\'a}s \cite{B03c} section 13, p25).
That is, the unusual slow scaling of the shortest path distance scale in this undirected version of this model is due to the effect of the very high degree nodes created because of the cumulative advantage (preferential attachment) process. \tnote{Is the Cohen and Havlin paper for BA model or generic scale-free graphs? See \cite{CH03}.}

However, the situation is completely different when we take account of the direction of edges in this model. First, the link between shortest path lengths and geodesics in Euclidean space used in \cite{EMY07,DMPP16} is lost. The natural order of nodes in a DAG, the arrow of time, means we should compare the path lengths network against geodesic lengths for network models embedded in Minkowski space, and indeed there is a proven relationship between these two \cite{BG91,RW09,CE16}. Following on from this, when using the longest path in the directed form of these models, our analysis has shown that the longest path is likely to be created by edges created from random attachment not those formed using the cumulative advantage mechanism, the opposite of what is suggested for the shortest path in the undirected form of these models.  Thus the fat-tailed nature of the degree distribution in the Price model (or directed versions of the Barab\'{a}si-Albert/LCD models) is not a factor for the longest path and so, using Bollob{\'a}s' insight \cite{B03c}, we should expect the longest path to scale simply as $\ln(t)$, and not something slower than that. That is, indeed what we have shown in our work here.

\tnote{In fact we conjecture that in these $m>1$ undirected form of these models, a shortest path of length equal to the diameter rarely passes through the first node. The early nodes are of high degree and likely to be part of a dense sub-graph. You are almost always going to be able to find shortcuts between nodes in this region so there is rarely any need to use more than one of them in a shortest path. There seems to be no simple relationship between the diameter found using shortest paths in the undirected case and the longest path in the DAG of the Price model.}

\tnote{For the special case of $m=1$, the model gives a tree.  In this case, in the undirected graph the shortest path is also the longest path between any two-vertices. However it is also true that the longest path length between any two vertices will never be between the first vertex and another vertex.  That is because the first vertex can be seen as being at the root of many branches.  We can find the length of any branch and this will be our longest path estimate in our DAG.  However, for the undirected case, we can also find the longest path along a second distinct branch. These two paths will be very similar in length, asymptotically the difference in the two paths will be a constant while the length of the paths grows as $\ln(t)$. In the undirected case, we may create a new path by combining these two paths along two distinct branches to give a longer path growing at twice the rate. This suggests that the longest path in the DAG version of the $m=1$ model grows at $(1/2)\ln(t)$, at half the rate of the diameter in the undirected case.}

Looking more widely, we note that Bollob{\'a}s \cite{B03c} (p10) suggested that ``For these models the orientation is not very interesting''. Our conclusion is the opposite. Namely that for any directed network in which vertices are added sequentially,
%including the Price model \cite{P76}, the Barab\'{a}si-Albert model \cite{BA99} model, the LCD model \cite{B03c,BR04},
the arrow-of-time inherent in these growing network models is both physically relevant and this vertex order produces new and distinctive features. Our analytical and numerical analysis of the longest path length is just one illustration of what is possible.

\section*{Acknowledgements}

V.V. acknowledges support from EPSRC, grant number EP-R512540-1.

\section*{Author contributions statement}

T.S.E.\ derived the analytical solution. L.C.\ and V.V.\ conducted the data analysis. T.S.E.\ and V.V.\ wrote the manuscript. All authors developed computational framework, analysed and interpreted the results and reviewed the manuscript.

\section*{Additional information}
%Data, supporting the findings, is available at Figshare DOI:10.6084/m9.figshare.7902584.
The authors declare no competing interests.

%%%\bibliography{longestpathprice}
% ***********************************************************************
%
% START BIBLIOGRAPHY

% END BIBLIOGRAPHY
%
% ***********************************************************************

\clearpage
% APPENDIX
%\begin{center}
%\large\textbf{Appendices}
%\end{center}
%\section*{\Large APPENDIX}
\section*{\Large Supplementary Information.}
%\appendix
%\renewcommand{\thesection}{A.\arabic{section}}
\renewcommand{\thesection}{\Alph{section}}
\setcounter{equation}{0}
\renewcommand{\theequation}{\thesection\arabic{equation}}
\renewcommand{\thefigure}{\thesection\arabic{figure}}
\renewcommand{\thetable}{\thesection\arabic{table}}
\setcounter{section}{0}

%%%\input{longestpathpriceapp}

% -----------------------------------------------------------------------------------------------------------
\section{Full Analytic Calculation}\label{aanalytic}

In this appendix we will work with a DAG $\Gcal$ with vertex set $\Vcal$ and edge set $\Ecal$.  The vertices will be labelled by sequential integers $t$ which take values from $1$ to the number of vertices $N=|\Vcal|$. The directed edges are defined to run from the lower value vertex to the higher value vertex, so if $(s,t) \in \Ecal$ then $s<t$.

% ..................................................
\subsection{Definition of reverse greedy paths}\label{agreedydef}

%The total order is given by the rank time of the nodes, i.e. the t-th node added to the network is node t.  The first node is the node at $t=1$ is the first one placed in the network, the only one with in-degree zero using the convention that edge go from node s to t where s<t (the direction of information flow if this is a citation network). (I may have switched convention in the Price model paper).   Of course in practice the initial starting graph is usually fixed in a different way but I assume there is always one of those designated t=1. That is the unique source vertex in the Price model whereas the last node added and several others are sink nodes in my conventions (out-degree zero).

Every vertex $t$ in a DAG $\Gcal$ has two reverse greedy paths: the \tdef{forward reverse greedy path} to nodes of times later than $t$, the other the \tdef{reverse greedy path} coming from nodes of earlier times. Here we will define the  reverse greedy path as that is what is used in this work.

The reverse greedy path arriving at a vertex $t$ is defined by
following the edge arriving at that vertex $t$ which links back to
the most recent predecessor vertex $s$. More formally, suppose
$\Npast(t)$ is the set of predecessors of a vertex $t$, the source
vertices of edges whose target vertex is $t$:
\begin{equation}
  \Npast(t) = \{ s | (s,t) \in \Ecal \}
\end{equation}
where $\Ecal$ is the edge set of the DAG. Then the reverse greedy
path to vertex $t$ is a sequence of vertices denoted as $\Ppast(t)$
which is defined to be
%\begin{widetext}
\begin{eqnarray}
\Ppast(t) &=& \{ s_i | s_i = \max(\Npast(s_{i+1})), \, i\in \{0, \ldots, \ellgr(t)\} , \,
\nonumber \\ && \qquad\qquad\qquad\qquad
s_0=1, \, s_{\ellgr(t)}=t \} \, .
\end{eqnarray}
%\end{widetext}
Note that the length of the reverse greedy path to
$t$ is denoted $\ellgr(t)$ so $\ellgr(t) = |P(t)|-1$. The reverse
greedy path will always terminate at a global \tdef{source node},
the only node with zero in-degree.

%Consider the construction of the following `reverse greedy path' from the unique source node (zero in-degree) at the initial time $t=1$ to any node $t$ back. We suppose that all earlier nodes know their reverse greedy path and its length $\ell(t)$.  The first step back on the reverse greedy path from node $t$ is made along the edge to the most recent successor node. The idea is that the most recent is also likely to be the one furthest from the  sink node, i.e.\ $\ell(t)$ is a monotonically increasing function of $t$.  So our reverse greedy path is $P_\mathrm{gr}$ where
%\bea
% P_\mathrm{gr} &=& \{ \tau_i | i \in \{ 0,1,\ldots,\ell(t) \}, \, \tau_\ell = t, \tau_0=1, \, (\tau_{i+1},\tau_i) \in \Ecal,
% \nnel
% &&
% \qquad\qquad\qquad\qquad\qquad
%  (\tau_{i+1}-\tau_i)\leq  (\tau_{i+1}-s) \, \forall \, (\tau_{i+1},s) \in \Ecal \}
%  \, .
%\eea

% -----------------------------------------------------

\subsection{Price Model Definition}\label{aprice}

%In the Price model  \cite{P65} (see sec.14.1  \cite{N09b}) we add one new vertex at time $t$.  This is connected to $m$ existing vertices $s$, each chosen with probability $\Pi(t,s)$. The node created at time $s$ will have in-degree $\kin(t,s)$ on average after the edges have been added to the node added at time $t$. This connection is made in one of two ways; with probability $p$ it is connected to an existing vertex $s$ chosen with cumulative advantage $\kin(t,s)/E(t)$ otherwise, so with probability $\pbar=(1-p)$, we choose the target vertex $t'$ uniformly at random, i.e. with probability $1/N(t)$.  That is

The Price model  \cite{P76} (see also sec.14.1 of  \cite{N09b})
is a growing network model giving us a sequence of graphs
$\{\Gcal(t)\}$ where these are indexed by an integer $t$, the
`time'. Each graph has a vertex set $\Vcal(t)$ and an edge set
$\Ecal(t)$. To create the next graph $\Gcal(t+1)$ in the Price
model sequence, we add one new vertex at time $(t+1)$ to
$\Vcal(t)$. We will overload our notation so that the positive
integer $t$ is both the `birth' time of a vertex and it is also
used to represent that vertex. So formally $\Vcal(t+1) = \Vcal(t)
\cup \{ (t+1)\}$. We choose to index our nodes from one upwards
so that at time $t$ the total number of nodes in the graph
$\Gcal(t)$, denoted as $N(t)$, is $N(t) = |\Vcal(t)| = t$.

A new node $(t+1)$ is connected by $m$ new edges to existing
vertices $s$ chosen with probability $\Pi(t,s)$. The number of
edges in the edge set $\Ecal(t)$ of the DAG $\Gcal(t)$, edges
between nodes created at time $t$ or earlier, is $E(t)=E_0 + mt =
|\Ecal(t)|$ where $E_0$ is some constant. Note this is the total
number of edges after we have finished adding all the incoming
edges to vertex, $t$.

Finally the node created at time $s$ in the graph $\Gcal(t)$ has out-degree $\kout(t,s)$.
% at the end of attachment to vertex $t$ so at the end of step $t$.

The connection of edges to new node $(t+1)$ as encoded by $\Pi(t,s)$ is made in one of two ways. With probability $p$ the node $(t+1)$ is connected to an existing vertex $s$ chosen with cumulative advantage $\kout(t,s)/E(t)$. Otherwise, so with probability $\pbar=(1-p)$, we choose an existing vertex $s$ uniformly at random from the set of existing vertices, i.e.\ with probability $1/N(t)$.  That is the probability of connecting new vertex $(t+1)$ to existing vertex $s$ is\footnote{In Newman's notation (equation (14.1), sec.14.1 of  \cite{N09b}) $\Pi(t,s)  = (\kout(t,s)+a)/(t(c+a))$ with  $a=m \pbar /p$ and $c=m$ (with $c+a=m/p$).}
\bea
  \Pi(t,s) &=&
  p \frac{\kout(t,s)}{E(t)} + \pbar \frac{1}{N(t)}
  \; \mbox{ if } t \geq s \geq  1  \mbox{ \& } t,s \neq 1  \, ,
  \nnel
  &=& 1 \mbox{ if } t=s=1 \, ,
  \nnel
  &=& 0 \mbox{ otherwise}
  \, .
  \label{aPricePidef}
\eea
The usual form, the first equation, leaves us, however, with a problem for $\Pi(t=1,s=1)$ when looking at the attachment to the second vertex, $t=2$, if we think of the initial vertex at $t=1$ as having no incoming edges. The solution is to insist that $\Pi(t=1,s=1)=1$. We will see that this fixes the cumulative probability $\Pi_\leq$ to have a consistent value which is in fact all we need for this calculation.

%\footnote{An alternative solution is to start the calculation from $t>1$ and probably for each of calculation with $E(t)=mt$ and $N(t)=t$ holding for $t \geq 1$. We would only need to allow for a non-zero initial path length $\ell(t=1)$ and this would give some control over initial conditions.}.

For simplicity we will almost always assume that $E(t)=mt$ and $N(t)=t$.  This is impossible to satisfy at early times as at the end of time step $1$ we have one node which cannot have the $m$ edges required since self-loops are not allowed for a DAG. In turn, this early time issue is a clear signal that there are some initial graph effects in the Price model. In practice, we can deal with this issue in various ways. If multiple edges between pairs of vertices are allowed then we can enforce this $E$ and $N$ constraint from step $t=2$ onwards (set $2m$ edges from node $1$ to node $2$) but at the cost of having a very skewed initial degree distribution. Without multiple edges, then the earliest point where we can have our $E(t)=mt$ and $N(t)=t$ is at $t=2m+1$ if at that time we have a complete directed graph in which the first $2m+1$ nodes have edges leaving them to connect to every later node, i.e.\ where the edge set is $\Ecal(t=2m+1) = \{ (i,j) | 1 \leq i < j \leq 2m+1\}$.
%\tnote{The degree distribution is reasonable while the distribution of shortest and longest paths is limited. In this directed complete graph, the shortest path is one for all but node $t=1$. The longest paths in the complete graph is $(t-1)$ for the node $t$. but perhaps no worse than any map of low integer values onto predicted average values.}
Unless otherwise stated, we used this complete graph graph at time $t=2m+1$ as the starting point for our numerical simulations.

We will need the cumulative probability function which is the probability that you attach all $m$ edges from the node created at time $(t+1)$ to a node created at time $s$ or earlier
\bea
 \Pi_\leq(t,s)
 &=& \sum_{r=1}^{s} \Pi(t,r) \; \; \mbox{if} \; \; t\geq s \geq  1 \, ,
 \\
 %\qquad \Pi_\leq(t,s)
 &=&0 \mbox{ otherwise.}
 \label{aPileqdef}
\eea
It is clear that we must have
\beq
 \Pi_\leq(t,s=t)=1 \, ,  \qquad t\geq 1
 \label{aPileqnorm}
\eeq
as all existing vertices are then of time $t$ or less. This shows that if $\Pi_\leq(t=1,s=1)=1$ then we need $\Pi_\leq(t=1,s=1)=\Pi(t=1,s=1)=1$ for consistency.

To find this cumulative advantage probability $\Pi_\leq(t,s)$ we
can work in terms of an effective node, a single `super' node,
which represents the $s$ nodes created from the initial time to
time $s$. If we coarse grain these nodes from $t=1$ to $s$ into one
super node, then we should attach all their edges to this one super
node (we can picture this as having multiple self-loops) so that
algebraically the degree of this super node, $\kouteff(t,s)$, is
the sum of all the degrees of the individual nodes created at or
before $s$, %\begin{widetext}
\bea
 \kouteff(t,s) &=& \sum_{r=1}^{s} \kout(t,r) \mbox{ if } t>s \, ,
 \nonumber \\
 \kouteff(s,s) &=& E(s) = m \, s \, .
  \label{akineffdef}
\eea
%\end{widetext}

Now we can use this to write the cumulative probability for attachment as
\bea
 \Pi_\leq(t,s)
 &=& \sum_{r=1}^{s}
  \left( p \frac{\kout(t,r)}{mt} + \pbar \frac{1}{t} \right)
\\
 &=&
  \frac{p}{mt} \left( \kouteff(t,s) + \frac{m \pbar}{p} s \right)
% =   \frac{p}{mt} y(t,s)
\eea

The master equation for this effective node is therefore
\bea
 \kouteff(t+1,s) -
 \kouteff(t,s)
 &=&
 \sum_{r=1}^{s} m\Pi(t,s)
 \nonumber
 \\
 &=&
 m\Pi_\leq (t,s)
% \\  &=&  \frac{p}{t} \left( \kouteff(t,s) + \frac{m \pbar s}{p} \right)
 \, .
\eea
We can rewrite this in terms of a function $y(t,s)$ where
\beq
 y(t,s) := \kouteff(t,s) + \frac{m \pbar s}{p} \, ,
 \mbox{  with  }
 y(s,s) = \frac{m s}{p}
 \, .
 \label{aydef}
\eeq
The master equation then becomes
\bea
 y(t+1,s) -
 y(t,s)
 &=&
 \frac{p}{t} y(t,s)
% \\ y(t+1,s)  &=&  \frac{p+t}{t} y(t,s)
 \\
\Ra \;\;\;
 y(t,s)
 &=&
 \frac{\Gamma(p+t)}{\Gamma(p+s)} \frac{\Gamma(s)}{\Gamma(t)}  \frac{m s}{p}
\eea
using $y(s,s)$ value in \eqref{aydef}.
This gives
\bea
\Pi_\leq(t,s)
 &=&
%  \frac{s}{t} \frac{\Gamma(p+t)}{\Gamma(p+s)} \frac{\Gamma(s)}{\Gamma(t)} =
  \frac{\Gamma(p+t)}{\Gamma(t+1)} \frac{\Gamma(s+1)}{\Gamma(p+s)}
  \, .
\eea
We note that this is separable with
%Note that in terms of the $h(t)$ and $j(s)$ functions for a separable cumulative advantage probability $\Pi_\leq(t,s)=j(s)/h(t)$, as defined in \eqref{aPileqsepdef}, we have here that
\beq
  \Pi_\leq(t,s) = \frac{j(s)}{j(t)} \, , \qquad
  j(s) = \frac{\Gamma(s+1)}{\Gamma(p+s)} \, .
  \label{jdef}
\eeq
For large $t$ and $s$ we have that
\bea
\Pi_\leq(t,s)
 & \approx &
 \left(\frac{s}{t}\right)^{\pbar}
 \, , \quad t,s \gg 1 \, .
%\\  &&   h(t) \approx t^{+\pbar} \, , \qquad   j(s) \approx s^{+\pbar} \, .
\eea

%Now we need to know the probability of connecting to the latest node at time $(t+1)$ and we are assuming that this will be the first connection on the longest path to the only global source node in this model, the initial node $t=1$ i.e.\ we are using a reverse greedy path calculation. The probability that we connect a new node $(t+1)$ to nodes of age $s$ or less is $(\Pi_\leq(t,s))^m$. So the probability that $s$ is the time of the latest node to which we connect new mode $(t+1)$, $\Pimax(t,s)$ is $\Pimax(t,s) =  (\Pi_\leq(t,s))^m -(\Pi_\leq(t,s-1))^m$ as given in \eqref{aPimaxdef}. We find that it is separable so we have that $\Pimax(t,s) =  J(s)/H(t)$ where
%\bea
% H(t) &=& \left( \frac{\Gamma(t+1)} {\Gamma(p+t)} \right)^m \, .
% \label{Hdef}
%\eea
%Later we will need the ratio
%\bea
% \Rightarrow \, \frac{H(t-1)}{H(t)}
% =
% \left( \frac{(s-\pbar)}{s}\right)^m \, .
% \label{Hratio}
%\eea
%\bea
% J(s) &=&
% \left( \frac{\Gamma(s+1)}{\Gamma(p+s)}  \right)^m
% -
% \left( \frac{\Gamma(s)}{\Gamma(p+s-1)}  \right)^m
% = H(s) - H(s-1)
%\eea

% ...................................................................
\subsection{Master Equation for Reverse Greedy Path Length $\ell$}\label{aspricemaster}

We are interested in looking at the reverse greedy path in our growing network model. For this we need to know $\Pimax(t,s)$, the probability that of the $m$ predecessor nodes connected to a new node at $(t+1)$, the oldest of them is $s =\max (\Npast(t+1))$.  The probability that we connect a new node $(t+1)$ to nodes of age $s$ or less is simply $(\Pi_\leq(t,s))^m$ where $\Pi_\leq(t,s)$ is the cumulative probability of attachment of \eqref{aPileqdef}.
This gives us that
\bea
\Pimax(t,s) &=&
 (\Pi_\leq(t,s))^m -(\Pi_\leq(t,s-1))^m
 \nonumber \\ && \qquad \qquad\qquad  \mbox{for  }
t\geq s \geq 1
 \, .
\label{aPimaxdef}
\eea
Note the case $s=1$ is covered as we defined $\Pi_\leq(t,0)=0$.
We may check that
\bea
\sum_{s=1}^t \Pimax(t,s) &=&
 (\Pi_\leq(t,t))^m
 =
 1
\eea
where we use the definition $\Pi_\leq(t,0)=0$ and \eqref{aPileqnorm}.

Let the probability that the length of the reverse greedy path, $\ell$, from new node $(t+1)$ to the initial node at $t=1$, be $P(\ell,t)$. This will satisfy the equation
\bea
 P(\ell,t+1)
 &=&
 \sum_{s=1}^{t} P(\ell-1,s) \Pimax(t,s) \, .
 \label{aGreedyMaster}
\eea
We can rewrite this master equation for $P(\ell,t)$ in terms of the generating function $G(z,t)$ defined as
\beq
 G(z,t) = \sum_{\ell=0}^\infty z^\ell P(\ell,t) \, ,
 \label{aGpricedef}
\eeq
where
\bea
G(z=1,t)
  &=& 1
  \label{aGzone}
   \\
G(z,t=1)
  &=& P(\ell=0,1)=1 \, .
\eea
This last follows because the only source node is the initial node at $t=1$ with reverse greedy path length zero,  $\ell(t=1)=0$.
%, so then we also have that $G(z,t=1)=P(\ell=0,1)=1$.
Ultimately, we want to look at the average length of the reverse greedy path, $\texpect{\ell(t)}$, which is obtained as
\bea
\left. \frac{\partial G(z=1,t) }{ \partial z }  \right|_{z=1}
  &=& \texpect{\ell(t)}
  \, .
  \label{aGtoell}
\eea

In terms of the generating function, the master equation \eqref{aGreedyMaster} becomes
\bea
 G(z,t+1)
 &=&
 \sum_{s=1}^{t} z G(z,s) \Pimax(t,s)  \, .
 \label{aGellmaster}
\eea
Peeling off the top term of the sum in \eqref{aGellmaster} gives us
\bea
 G(z,t+1)
 &=&
 z G(z,t) \Pimax(t,t)
 \nonumber \\ &&
 +
 \sum_{s=1}^{t-1} z G(z,s) \Pimax(t,s)  \, .
 \label{aGellmaster2}
\eea
For now, we will simply assume that cumulative probability $\Pi_\leq(t,s)$ is separable, i.e.\
$\Pi_\leq(t,s) = h(t) j(s)$
for some functions $h(t)$ and $j(s)$. We can use the fact that $\Pi_\leq(t,s=t)=1$ from \eqref{aPileqnorm} to see that $h(t) = 1/ j(t)$ so that $\Pi_\leq(t,s) = {j(s)}/{j(t)}$.  We have the precise form for $j(s)$ in the Price model in \eqref{jdef}.
That gives us that $\Pimax(t,s)$ is also separable and this can be written as
\bea
\Pimax(t,s)
 &:=&
 %\frac{J(s)}{H(t)}=
% \left(\frac{j(s)  }{j(t)}\right)^m
%   -\left(\frac{j(s-1)}{j(t)}\right)^m
% =
 \frac{H(s)-H(s-1)}{H(t)}
\label{aPimaxsepdef}
\\
H(t) &=& (j(t))^{m} \, .
\label{aHdef}
\eea
Then
%\begin{widetext}
\bea
 \lefteqn{G(z,t+1)
 =
 z G(z,t) \Pimax(t,t) }\shoveleft
 \nonumber \\ && \quad \quad
 +
 \sum_{s=1}^{t-1} z G(z,s)  \frac{ H(s)-H(s-1)}{H(t)}
%\\ &=& z G(z,t) \Pimax(t,t) + \frac{H(t-1)}{H(t)} \sum_{s=1}^{t-1} z G(z,s)  \frac{ H(s)-H(s-1) }{H(t-1)}
\\
 &=&
 z G(z,t) \Pimax(t,t)
 \nonumber \\ &&
 +
 \frac{H(t-1)}{H(t)} \sum_{s=1}^{t-1} z G(z,s)  \Pimax(t-1,s)  \, .
\eea
%\end{widetext}

Using \eqref{aGellmaster} leaves us with
\bea
 G(z,t+1)
 &=&
 \left( z \Pimax(t,t) +  \frac{H(t-1)}{H(t)}  \right)G(z,t)
 \quad {}
 \label{aGellmaster3}
%\\ &=&  \left( z \frac{H(t) - H(t-1)}{H(t)}  +  \frac{H(t-1)}{H(t)}  \right)G(z,t)  \, ,  \label{aGellmaster3b}
\\
 &=&
 \left( z   + (1-z) \frac{H(t-1)}{H(t)}  \right)G(z,t)  \, ,
 \label{aGellmaster3b}
\eea
which has the formal solution
\bea
 G(z,t)
 &=&
 \prod_{s=1}^{t-1} \left( z +  (1-z) \frac{H(s-1)}{H(s)}  \right)
 \label{aGellmaster4}
\eea
given $G(z,1)=1$.  Now we can check $z=1$ value which is clearly $G(z=1,t)=1$ as required.

We can obtain the precise form for the Price model by using the form for $H(s)$ in \eqref{aHdef} given $j(s)$ in the Price model from \eqref{jdef}.  This gives us that
\beq
\frac{H(s-1)}{H(s)} =  \left( \frac{s-\pbar}{s} \right)^m
\, .
\label{aHprice}
\eeq
The solution for the generating function then becomes
\bea
 G(z,t)
 &=&
 \prod_{s=1}^{t-1} \left( z \left( 1  - \left( \frac{s-\pbar}{s}  \right)^m\right)
  +
  \left( \frac{(s-\pbar)}{s} \right)^m
  \right)
  \, .
  \nonumber \\ && {}
  \label{aGsol}
\eea
In principle this is a ratio of $m$-th order polynomials in $s$ so you can write this product as the product of $m$ ratios of Gamma functions, one term per root of the polynomial.

% ...................................................
\subsubsection*{Average Path Length}

We can find $\ell(t)$, the average length of the reverse greedy path in the Price model, from this general solution
\eqref{aGellmaster3} using \eqref{aGtoell} and $G(z=1,t)$ from
\eqref{aGzone}.  This gives us that
%\begin{widetext}
\bea
\frac{\partial }{ \partial z}   G(z,t)
 &=&
 \sum_{s=1}^{t-1}\left[
 \left( 1- \frac{H(s-1)}{H(s)}\right)
 \prod_{r=1, r\neq s}^{t-1} \left( z +  (1-z) \frac{H(r-1)}{H(r)}  \right)
 \right]
\\
 &=&
 \sum_{s=1}^{t-1}\left[
 \left( \frac{1- (H(s-1) / H(s) ) }{z +  (1-z) (H(s-1)/H(s))}
 \right)
 \prod_{r=1}^{t-1} \left( z +  (1-z) \frac{H(r-1)}{H(r)}  \right)
 \right]
\\
 &=&
 \left[
 \sum_{s=1}^{t-1}
 \left( \frac{H(s)- H(s-1) }{z H(s) +  (1-z) H(s-1) }
 \right)\right]
 G(z,t)
\eea
%\end{widetext}
so then
\bea
\texpect{\ell(t)} &=&
 \left. \frac{\partial }{ \partial z}   G(z,t) \right|_{z=1}
  \\
 &=&
 \left[
 \sum_{s=1}^{t-1}
 \left( \frac{H(s)- H(s-1) }{H(s) }
 \right)\right]
 G(z=1,t)
 \quad {}
 \label{aellsol}
\eea
and finally
\bea
\texpect{\ell(t)} &=&
% \sum_{s=1}^{t-1}  \left( \frac{H(s)- H(s-1) }{H(s) }  \right) =
 \sum_{s=1}^{t-1}
 \left( 1-\frac{H(s-1) }{H(s) }
 \right)
 \label{aellsol3}
\eea
For the Price model we have $H(s)$ from \eqref{aHprice} and so we find that
\bea
\texpect{\ell(t)} &=&
  \sum_{s=1}^{t-1}
  \left( 1  - \left( \frac{s-\pbar}{s}  \right)^m\right)
  \label{aellsolgen}
\eea
Now we can expand the product using the binomial expansion to find that
\bea
\texpect{\ell(t)} &=&
  \sum_{n=1}^{m} \binom{m}{n}
  (-1)^{n-1} (\pbar)^n
  \sum_{s=1}^{t-1}s^{-n}
  \label{aellsolgen2}
\eea

For each value of $n$ we can write the $s^{-n}$ series as the difference of two Hurwitz zeta functions $\zeta(m,t)$
%, and in turn in terms of the polygamma functions $\psi^{(n-1)}(t)$, as follows
\bea
\texpect{\ell(t)} &=&
  \sum_{n=1}^{m} \binom{m}{n}
  (-1)^{n-1} (\pbar)^n \left( \zeta(n,1)-\zeta(n,t)\right) \, .
  \quad \quad {}
  \label{aellsolgen3}
%  \\
%  \zeta(n,t) &:=& \sum_{i=0}^{\infty} \frac{1}{(i+t)^n} = \frac{(-1)^{n}}{(n-1)!} \psi^{(n-1)}(t)
%  = \frac{(-1)^{n}}{(n-1)!} \frac{d^{n-1}}{dt^{n-1}} \ln(\Gamma(t))
\eea
The Hurwitz zeta functions are finite for $t>0$ and $n>1$ so we see that the terms $n\geq2$ only contribute a constant plus terms from $\zeta(n,t)$ which fall off as $t^{1-n}$ or faster.

The leading term in the large time limit of $\texpect{\ell(t)}$ comes only from the $n=1$ term in \eqref{aellsolgen2} and this is
\bea
\texpect{\ell(t)}_{n=1}
%  &\approx& m \pbar \left(H_{t-1} - H_{1-1} \right) + O(\pbar^2)   \label{aellasymp1} \\
 &\approx&
 m\pbar \ln(t) - m\pbar \psi(m\pbar + 1)
 + O(t^{-1}) \label{aellasymp2}
 %- \frac{m\pbar (1/2)-m\pbar}{t}   + O(t^{-2}) \label{aellasymp2}
  \quad {}
\eea
where $\psi(t)$ is the digamma function.

The constant in an asymptotic expansion in $t$ of
$\texpect{\ell(t)}$ picks up further contributions from the $n\geq
2$ terms so we have that
%\begin{widetext}
\bea
\lim_{t \ra \infty} \texpect{\ell(t)} &=&
   m\pbar \ln(t) - m\pbar \psi(m\pbar + 1)
  \nonumber \\
  &+&
  \sum_{n=2}^{m} \binom{m}{n}
  (-1)^{n-1} (\pbar)^n \zeta(n)
 + O(t^{-1}) \label{aellasymp2b}
 %- \frac{m\pbar (1/2)-m\pbar}{t}   + O(t^{-2})
   \quad \quad {}
   \label{aellasymp2c}
\eea
%\end{widetext}
where it is implicit that there is no contribution
from the term with the sum for the case of $m=1$. Here $\zeta(n)$
is the Riemann zeta-function.

% ..................................................
% Variance

% ...................................................
\subsubsection*{Variance}

We can also use the generating function solution \eqref{aGsol} to find the variance $\sigma^2(t)$ of the reverse greedy path length. We have that
\begin{eqnarray}
\texpect{\ell(t)(\ell(t)-1) }
 &=&
 \left. \frac{\partial^2 }{ \partial z^2}   G(z,t) \right|_{z=1}
\end{eqnarray}
We have that
\begin{eqnarray}
 \frac{\partial^2 }{ \partial z^2}   G(z,t)
 &=&
 \left[
 \left(
 \sum_{s=1}^{t-1}
 \frac{H(s)- H(s-1) }{z H(s) +  (1-z) H(s-1) }
 \right)^2
 \right.
 \nonumber \\
 &&
 \qquad \qquad
 \left.
 -
 \Bigg(
 \sum_{s=1}^{t-1}
 \left( \frac{(H(s)- H(s-1))^2 }{(z H(s) +  (1-z) H(s-1) )^2 }  \right)
 \Bigg)
 \right]
 G(z,t)
\end{eqnarray}
so that
\begin{eqnarray}
\texpect{\ell(t)(\ell(t)-1) }
 &=&
 \left[
 \left(
 \sum_{s=1}^{t-1}
 \frac{H(s)- H(s-1) }{H(s)}
 \right)^2
 -
 \Bigg(
 \sum_{s=1}^{t-1}
 \left( \frac{(H(s)- H(s-1))^2 }{( H(s) )^2 }  \right)
 \Bigg)
 \right]
 \\
 &=&
 (\texpect{\ell})^2
 -
 \Bigg(
 \sum_{s=1}^{t-1}
 \left( \frac{(H(s)- H(s-1))^2 }{( H(s) )^2 }  \right)
 \Bigg)
\end{eqnarray}
With $\sigma^2(t) = \texpect{(\ell(t))^2} - (\texpect{\ell(t)})^2$ we have that
\begin{eqnarray}
 \sigma^2(t) - \texpect{\ell(t)}
 &=&
 \texpect{\ell(t)(\ell(t)-1) } - (\texpect{\ell(t)})^2
 \\
 &=&
 -
 \sum_{s=1}^{t-1}
 \left( 1 -  \frac{H(s-1)}{H(s)}    \right)^{2}
 \\
 &=&
 -
 \sum_{s=1}^{t-1}
 \left( 1 - \left(\frac{(s-\pbar)}{s} \right)^{m}   \right)^2
 \\
 &=&
 - (m \pbar)^{2}
 \sum_{s=1}^{t-1}
 (s)^{-2}
 + O(m^3\pbar^3)
\end{eqnarray}
In the long-time limit we have that
\begin{eqnarray}
 \lim_{t\ra \infty}
 \sigma^2(t) - \texpect{\ell(t)}
 &=&
 - \frac{\pi^2}{6} m^2 (\pbar)^2
 + O(m^3\pbar^3)
\end{eqnarray}
Note that these correction are all finite so we have basically shown that the variance grows with the mean i.e.\ Poisson-like behaviour.

% ...................................................
\subsubsection*{Long-time Generating Function Result}

From \eqref{aGsol} the logarithm of the generating function $G(z,t)$
\bea
 \ln(G(z,t))
 &=&
 \sum_{s=1}^{t-1}
 \ln \left(1
      + (z-1) \left( 1- \left(1-  \frac{\pbar}{s} \right)^m   \right) \right) \, .
  \label{aGln}
\eea
%Using the binomial expansion this can be written as
%\bea
% \ln(G(z,t))
% &=&
% \sum_{s=1}^{t-1}
% \ln \left(
%      1+
%        (1-z) \sum_{n=1}^{m} \binom{m}{n}  (-1)^{n} (\pbar/s)^n
%  \right) \, .
%  \label{aGln2}
%\eea
Expanding around $z=1$, or with $\pbar$ small, we have that
\bea
 \ln(G(z,t))
 &=&
 -
 \sum_{q=1}^\infty
 \frac{(1-z)^q}{q}
 \sum_{s=1}^{t-1}
 \left( 1- \left(1-  \frac{\pbar}{s} \right)^m   \right)^{q}
  \, .
  \label{aGln3}
\eea
We recognise the first term as $\texpect{\ell(t)}$ from \eqref{aellsolgen} so that
\bea
 \ln(G(z,t))
 &=&
 (z-1) \texpect{\ell(t)}
 -
 \frac{(1-z)^2}{2}
 \sum_{s=1}^{t-1}
 \left( \frac{(m\pbar)^2}{s^2}   +\ldots  + \frac{\pbar^{2m}}{s^{2m}} \right)
 \nonumber \\
 &&
 \qquad \qquad \qquad\qquad
 -
 \sum_{q=3}^\infty
 \frac{(1-z)^q}{q}
 \sum_{s=1}^{t-1}
 \left( 1- \left(1-  \frac{\pbar}{s} \right)^m   \right)^{q}
  \label{aGln4}
  \\
 &=&
 (z-1) \texpect{\ell(t)}
 -
 \frac{(1-z)^2}{2}
 \pbar^2 \left( m^2\zeta(2)  +\ldots  + \zeta(2m) \pbar^{2(m-1)} \right)
 \nonumber \\
 &&
 \qquad \qquad \qquad\qquad -
 O(\pbar^3(1-z)^3)\, .
  \label{aGln5}
\eea
That is the leading term comes from the only term with a sum over $1/s$ which diverges as $\ln(t)$ and is completely captured by the $\texpect{\ell(t)}$ contribution. The higher order terms in an expansion around $z=1$ are all finite which may be expressed as polynomials in $\pbar$ with terms from $\pbar^q$ to $\pbar^{qm}$ for the coefficient of the $(1-z)^q$ term.

From this we wee that the leading order term in $G(z,t)$ in the long time limit
\bea
 \lim_{t \ra \infty} G(z,t))
 &=&
 \exp( (z-1) \texpect{\ell(t)} )
\label{aGlimt}
\eea
which is the generating function for a Poisson distribution.  So we deduce that in the long-time limit, the distribution of lengths of the reverse greedy path in the Price model is a Poisson distribution with mean equal to $m\pbar \ln(t)$.
\beq
 \lim_{t \ra \infty} P(\ell,t) = \frac{e^{-\lambda} \lambda^ \ell}{\ell!} \, \qquad
 \lambda = \texpect{\ell(t)} \approx m\pbar \ln(t)
 \, .
\eeq
% ..................................................
\subsection{Reverse Greedy Path for $m=1$}

The Price model for $m=1$ is a special case in this model as then the graphs produced are directed trees. Various aspects of this $m=1$ case of the Price model and its variants can be studied analytically, for example  \cite{CJ13,HIH17}. Here we just note that many of the equations in our calculations become simple first order polynomials, such as for $\Pimax(t,s)$ in \eqref{aPimaxdef}, which allows for a simple direct solution in the $m=1$ that case. The solutions for the reverse greedy path length also takes on a much simpler form.

For $m=1$ we have from \eqref{aGsol} that the generating function is just
\bea
 G(z,t;m=1)
 &=&
 \prod_{s=1}^{t-1} \frac{(s+(z-1)\pbar)}{s}
  =
  \frac{\Gamma(t+(z-1)\pbar)}{\Gamma(1+(z-1)\pbar) \Gamma(t)}
  \, .
  \label{aGm1sol}
\eea
\tnote{The long time limit is not quite the generating function of the Poisson distribution --- I find $G(z,t,m=1) \sim \exp((z-1)\pbar\ln(t)) / \Gamma(1+(z-1)\pbar)$.}
%\bea
% \lim_{t \ra \infty} G(z,t;m=1)
% &=&
%  \frac{\exp((z-1)\pbar\ln(t))}{\Gamma(1+(z-1)\pbar)}
%  \, .
%  \label{aGm1soltinf}
%\eea

Similarly, the expression for the reverse greedy path length is also simple:
\bea
\texpect{\ell(t)} &=&
  \sum_{s=1}^{t-1}
  \frac{\pbar}{s}
  =
  \pbar (\psi(t) - \gamma)
  \label{aellm1sol}
\eea
where $\gamma\approx 0.577 $ is the Euler-Mascheroni constant.

%For $m=2$ we no special features but the forms become quadratic and still relatively tractable.  They illustrate the extra terms arising for $m>1$ solutions.
%From \eqref{aGsol}, the generating function  solution is
%\bea
% G(z,t)
% &=&
% \prod_{s=1}^{t-1} \left( z \left( 1  - \left( \frac{s-\pbar}{s}  \right)^m\right)
%  +
%  \left( \frac{(s-\pbar)}{s} \right)^m
%  \right)
%  \, .
%  \label{aGsol}
%\eea

% ..................................................
\subsection{Vertex Partitions}\label{avpart}

% \tcomment{
% \begin{itemize}
% \item Mention link to antichains
% \item Do approximate calculations of size, mean time (index) and variance of time.
% \item Do for reverse greedy path partition and longest path partition.
% \item Do for $\pbar m \gtrsim 2$ and $m\geq 3$ and may be two or three parameter values.
% \end{itemize}
% }

%The reverse greedy path partition $\Pcal(\ellgr)$ is defined in terms of the reverse greedy path length $\ellgr$)
%\begin{equation}
%\Pcal(\ellgr) = \{ t \, | \, t \in \Vcal , \, \ellgr(t) = \ellgr \} \, .
%\label{agreedyPdef}
%\end{equation}

A partition $\Pcal$ of the vertex set $\Vcal$ is a set of
non-overlapping non-empty subsets which contain each and every
vertex once and only once. That is $\Pcal = \{ \Pcal(i) \}$ where
the
\tdef{blocks} of the partition, $\Pcal(i)$, are such that $\Pcal(i)
\subset \Vcal$, $\Pcal(i) \neq \emptyset$, $\Pcal(i) \cap \Pcal(j) = \emptyset$ unless $i=j$,
and $\cup_i \Pcal(i) = \Vcal$. The definition of a unique integer
length scale associated with each node in any one instance of the
model gives a natural partition of the set of vertices.
%For example, one partition in any DAG is based on the length of the
reverse greedy path to each node $t$ from a source node,
$\ellgr(t)$.

Of particular interest here is a partition in terms of the longest
path $\ellmax(t)$ to a given node $t$. In any DAG, the
\tdef{height} of a node is the length of the longest path to a node
from any global source node, any node with zero in-degree. Thus in
the Price model, the height of a node is simply the longest path
length from the initial node to any node, that is our $\ellmax(t)$
is the height. A natural partition of the set of vertices in a DAG
is where each block, $\Pcalh(h)$, contains all the vertices of
height $h$,
\begin{equation}
\Pcalh(h)
 =
 \{ t \, | \, t \in \Vcal , \, \ellmax(t)=h \}
 \, ,
 \quad h \in \Zbb
 \, .
\label{aPhdef}
\end{equation}
This partition by height has the special property that no two vertices in any one block are connected by any path. In a DAG nodes connected by a path cannot be of the same height. A set of disconnected vertices is known as an \tdef{antichain} so the blocks of the height partition are all antichains. In some sense, the vertices in an antichain can be considered to be of equivalent ages. There is no relationship between them which says that any one vertex in the antichain need to come before or after another.
The question here is can we estimate some of the properties of these height antichains? To do that we need to define a slightly different partition.

For the Price model, we can also define another type of partition in terms of the average path length $\texpect{\ellgr(t)}$ associated with a node created at time $t$ averaged over all possible instances of the Price model. We will discuss this partition in terms of the reverse greedy path length $\ellgr$ where we have analytical results but the construction is identical for the longest path length $\ellmax$.
Since our average path lengths are monotonically increasing functions of time $t$, we can define the blocks $\Pcalhat(\ellgrhat)$ of one partition in terms of an integer value $\ellgrhat \in \Zbb$ as follows. All nodes in one block of the partition have an expectation value for their reverse greedy path length which rounds to the integer $\ellgrhat$,
\begin{equation}
\Pcalhat(\ellgrhat)
 =
 \{ t \, | \, t \in \Vcal , \, \ellgrhat-\half \leq \texpect{\ell(t)} < \ellgrhat+\half , \, \ellgrhat \in \Zbb^+\} \, .
\label{aPhatdef}
\end{equation}
%From our analysis we have that our length scales
%\bea
%\texpect{\ell (t)}
%  &\approx&
% m\pbar \ln(t)
% \label{aellgrapprox}
%\eea
%so that for large times

The idea here is that we can estimate the properties of the blocks in this partition, $\Pcalhat(\ellgrhat)$, by using the leading behaviour for the lengths scales, namely they grow logarithmically with $t$. We would then expect that the partitions of a single instance based on the measured values of an integer valued path length would show similar features.

The leading behaviour for the mean reverse greedy path length of a node $t$ can be expressed as
\begin{equation}
 t = \tilde{t} (\mugr)^{\ellgrbar}
 \label{aellgrapprox}
\end{equation}
where $\ellgrbar = \texpect{\ell(t)}$ while $\mugr$ and $\tilde{t}$ are some constants.  Our leading order results give $\mugr=\exp(1/(m\pbar))$ for the reverse greedy path, while we can estimate $\mu = \exp(1/a)$ and $\tilde{t}=\exp(-b/a)$ using the numerical results for the coefficients $a$ and $b$ obtained from the fit to $a\ln(t)+b$.

From this behaviour we can estimate various properties of the blocks $\Pcalhat(\ellgrhat)$. It is useful to express these results in terms of the time scale $\hat{t}(\ellgrhat)$ (not necessarily an integer) for a node whose path length is expected to be the integer $\ellgrhat$, namely
$\hat{t}(\ellgrhat) = \tilde{t} (\mugr)^{\ellgrhat} $ for $\ellgrhat \in \Zbb^+$.  Other properties of the path length partition $\Pcalhat(\ellgrhat)$ can be estimated in terms of $\hat{t}$: the average time of nodes in the partition $\tbar(\ellgrhat)$, the standard deviation of the times of nodes in the partition $\sigma_t(\ellgrhat)$ and the number of nodes in the block $ |\Pcalhat|(\ellgrhat)$.  We find that
\begin{eqnarray}
\tbar(\ellgrbar)
 &=&
 \cosh(\sqrt{\mugr})  \, \hat{t}(\ellgrbar)
 \label{atbargr}
\\
\sigma_t(\ellgrbar)
 & \approx &
 \frac{1}{\sqrt{3}} \sinh(\sqrt{\mugr}) \, \hat{t}(\ellgrbar)
 \label{asigmagr}
\\
 |\Pcalhat|(\ellgrbar)
 &=&
 2\sinh(\sqrt{\mugr}) \,  \hat{t}(\ellgrbar)
 \label{apartpropgr}
\end{eqnarray}

The results above are for the partition into blocks $\Pcalhat(\ellgrhat)$ of \eqref{aPhatdef} which are defined in terms of averages over an ensemble.  More interesting is to look at the partition into height antichains, $\Pcalh(h)$ of \eqref{aPhdef} since this based on the topology of a single network and so much more relevant to studies of real data sets. Our suggestion is that in the Price model all the length scales behave in the same way, following our description in \eqref{aellgrapprox}.  So we should expect that all partitions in the Price model based on such length scales should show the scaling behaviour suggested in \eqref{apartpropgr}.  For example that suggests that for the height antichains, we should compare the ratios of a particular property for adjacent blocks, for instance $\tbar(h+1)/\tbar(h)$, since this analysis suggests such ratios will be constant.

\section{Numerical Implementation}\label{anumeric}

% --------------------------------------------------------------
\subsection{Price Model Algorithm}\label{apricealg}

To reduce the number of random numbers drawn, we use the following
algorithm. First we observe that the probability $\Pi(t,s)$ of
\eqref{aPricePidef} used to choose existing nodes $s$ to attach to
a new node $(t+1)$ is proportional to $\Pi(t,s) \propto \kout(t,s)
+ (m\pbar/p)$. If we limit ourselves to parameter values where
$\alpha = (m\pbar/p)$ is a non-negative integer, then we can choose
existing vertices with probability $\Pi(t,s)$ by drawing uniformly
at random from a list, \verb$attachment_list$, if it is properly
formed. We define our algorithm as follows.
\begin{enumerate}
\item Set initial graph \verb$G$ with $t$ nodes.
\item Initialise \verb$attachment_list$ to match initial graph.
\begin{itemize}
\item For every edge in the initial graph, we add the source end of the edge to  \verb$attachment_list$. That is is $(s,t) \in \Ecal(t)$ where $s<t$ then we add $s$ only to \verb$attachment_list$.
\item Each node in the initial graph is added $\alpha=(m\pbar/p)$ times to  \verb$attachment_list$.
\end{itemize}
\item  \label{aloop2} Increment $t$.
\item Draw $m$ times uniformly at random from \verb$attachment_list$, to give the list (sequence) $S(t) = [s_1, s_2, \ldots,s_m]$.
\begin{itemize}
\item If drawn with replacement, this is simple and fast and it matches the assumptions implicit in the algebraic calculations.  This allows multiple edges between nodes.
\item If drawn without replacement we avoid multiple edges but this is liable to be slower.  However this constraint is not enforced in the analytic equations.
\end{itemize}
\item Append the $m$ nodes in the list $S(t)$ to \verb$attachment_list$.
\\
This captures the cumulative advantage process.
\item Append $\alpha=(m\pbar/p)$ copies of $t$ to \verb$attachment_list$.
\\
This encodes the uniform random process.
\item If adding more nodes, return to step \ref{aloop2}.
\end{enumerate}
We allowed multiple edges in our simulations and in this approach one random number was needed for every edge added, a total of $mN$ random numbers if the final network had $N$ nodes.  For the attachment list $(m+\alpha)N$ memory locations are needed. Note that in our work here we only created the \verb$attachment_list$ along with lists for the length of paths to each vertex.  There was no need for use to create the full network structure. However, should the full network structure be required, it can be deduced later from \verb$attachment_list$ as defined here.

Note that the $p=0$ case is treated separately since the attachment list is not needed. In that case we draw uniformally at random from the set of nodes, with replacement in our case since we allow multiple edges.

In our work we choose to allow multiple edges in the networks created in our simulations.  This will have relatively little effect on results as the model is inherently a sparse graph model.  As the number of nodes increases, as $m$ is fixed, the chance of choosing the same existing vertex twice when connecting it to a new node decreases.

As noted elsewhere, our initial graph was a DAG of $2m+1$ nodes with an edge between every node, that is the initial edge set was $\Ecal(2m+1) = \{ (s,t) | 1\leq s < t \leq 2m+1 \}$. This has $E(t)=mt$ for $t\geq 2m+1$. However, we did investigate the effect of the initial graph and this is discussed in \secref{ainit}.

An alternative algorithm for the Price model, which we did not use here, would be to carry two separate lists of existing vertices, one used for the cumulative advantage process, and one for the selecting uniformly at random  from the set of vertices.  This is equally easy to code. The disadvantage is that we would now need to draw an additional random number for every edge added, a total now of $2mN$, as this extra random number has to be compared against the parameter $p$ in order to decide which list to draw from. The extra random numbers would slow the process but the gain would be flexibility as there would be no restriction on the parameters to ensure that $\alpha=(m\pbar/p)$ was integer. This alternative method would also only require $(m+1)N$ memory locations for the lists.

% --------------------------------------------------------------
\subsection{Numerical Reverse Greedy Path Algorithm}

Numerically, the reverse greedy path may be found as follows
\begin{enumerate}
\item Set initial graph \verb$G$ with $t$ nodes.
\item Set the reverse greedy path length for the initial $t$ nodes, \verb$greedy_length[s]$ for $s = 1,2,\ldots,t$.
\item Set the nearest neighbour along this reverse greedy path length for the initial $t$ nodes, \verb$greedy_neighbour[s]$ for $s = 1,2,\ldots,t$.
%\\
%Note that the %\verb$dag.node[1][greedy_neighbour]=None$,
%\verb$greedy_neighbour[1]=None$,
%\\
%%\verb$dag.node[1][greedy_length] = 0$
%\verb$greedy_length[1] = 0$
%Numerically  if the first node t=1 is indexed as node zero (I don't think you need this shift in python but its convenient in python and C++)

\item  \label{aloop} Increment $t$.
\item  Add a new node index $t$ to \verb$G$.
\item Update graph \verb$G$ by adding $m$ new edges, from vertices $\{s\}$ to new node $t$, which then define the past neighbour set of vertex $t$, i.e.\ $ \Npast = \{ s \} $.
\item Find the past neighbour with the largest time $s_\labelgr = \max ( \Npast(t) )$.
\item Record this closest past neighbour as the last step on the reverse greedy path to $t$,
\\
i.e.\ set
%\verb$dag.node[t][greedy_neighbour] = max (dag.predecessors(t)) = smax$.
\verb$greedy_neighbour[t]$ $=s_\labelgr$.
\item Set the length of the reverse greedy path of new node $t$ to be one more than the length of the reverse greedy path to this nearest past neighbour
    $s_\labelgr $, so set $\ellgr(t) = \ellgr( s_\labelgr   )+1$.
Numerically we just need to set
\\
%\verb$dag.node[t][greedy_length]= dag.node[smax][greedy_length]$.
\
\verb$greedy_length[t]=$

\verb$greedy_length[greedy_neighbour[t]]+1$.
\
\item If adding more nodes, return to step \ref{aloop}.
\end{enumerate}

The length of the reverse greedy path from the first node $t=1$ to a given node $t$ is now stored for each node.  The vertices on the unique reverse greedy path can be found by iterating back though the \verb$greedy_neighbour$ property of the nodes.  If the specific path used is not required we need not record this information.
%This means we need only store the reverse greedy path length $\ellgr(s)$ for each vertex
%e.g.\ as  \verb$dag.node[s][greedy_length] = \ellgr(s)$.
%If you want to be able to recall the reverse greedy path, you need only record the previous vertex on this reverse path e.g.\ \verb$dag.node[s][greedy_neighbour]$.

We also note that the length of the longest path, the longest path from the initial node $t=1$ to any node $t$, may be tracked numerically in a similar way.  Such a longest path is always exists but is not unique so we could only ever record one example longest path using this approach.

% ..................................................
\subsection{Fitting}\label{afit}

For each set of parameters $p,m, N$, we ran the model $R=100$ times.\vnote{is N a variable if we only consider the largest possible value we were able to handle, i.e. $10^8$?} We collected the length of the reverse greedy path and of the longest path for each node created at time $t$. In this case, each data point $y_{tr}$ is the relevant path length, $\ell(t)$ or $L(t)$, measured for node $t$ on run $r$.

\begin{figure*}[!htb]
  \includegraphics[width=\textwidth]{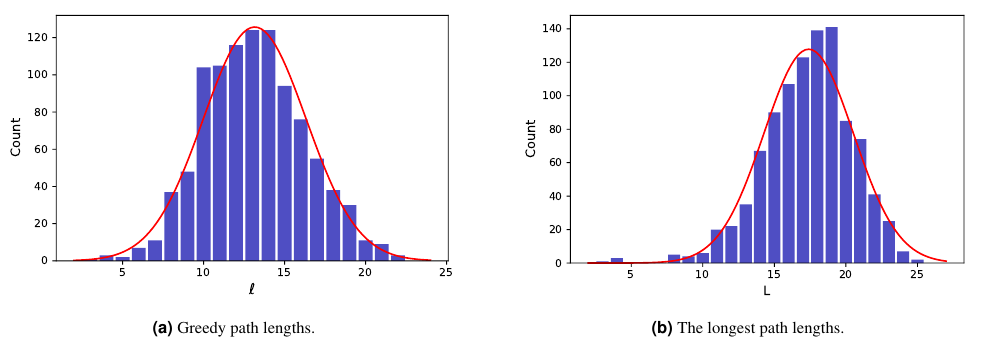}
\caption[Path length distributions]
    {\small Distributions of the reverse greedy path length (left) and longest path lengths (right) for a node with index $t=10^5$ in the Price model with $m=5$, $c=1$ and $10^6$ total number of nodes, averaged over 1000 networks. For comparison, a Gaussian is plotted with the same mean and standard deviation as each data set. We can see that a normal distribution,, while not perfect, is a sufficiently good description of the distribution to justify our use of gaussian based fitting statistics.}
%%%    {\small Distributions of the longest (top left) and greedy (top right) path lengths for a node with index $t=1000$ in the Price model with $m=5$, $c=1$ and $10^6$ total number of nodes, averaged over 1000 networks. For comparison, bottom left figure and bottom right figure show relevant Gaussian-distributed random values with similar mean and standard deviation. Visually, the lengths of longest paths seems to be reasonably well described by a normal distribution, but the distribution of longest path lengths at this early time shows ???. .}
    %\caption{Distribution in greedy (left) and longest (right) path lengths for node $t=???$ over ??? runs of $p=???$ $m=3$ or similar.\tcomment{Version of Fig.8 of  \cite{C19}.}}
\label{fonerunfluct}
\end{figure*}

We make several assumptions about our data. First, we assume that the measurements we make on each individual run are independent.
%That means, for instance, that if we measure a high value at point $t$ we are not more likely to measure a low value at $t+1$.
This assumption is not completely true in our data. \tnote{I've been trying to explain in one sentence why but can't think of exactly why but sure it is true.}
\vnote{? as the longest path to any node is bound to be larger than that to any of its neighbours by no more than one edge  ?}
Secondly, we assume that the set of measurements at the same point $t$ across many runs have a distribution that is normally distributed. As \figref{fonerunfluct} shows, this is approximately true.
\tnote{This next bit is not about what we did so do we need it? Comment it out perhaps? We may also conduct standard tests of normality, such as Shapiro-Wilk test \cite{SW65}, Anderson-Darling test \cite{AD54}, D`Agostino's K-squared test \cite{RD70} to name just a few. They are, however, likely to reject our hypothesis that the data follows a Gaussian distribution, as our data is discrete.}
\vnote{happy to delete them. I did conduct the tests and most of them failed, both for our discrete data and discretised normal distribution.}
%The normal distribution at each point $t$ means is that we can take the mean $\bar{y}_t$ and standard deviation $\sigma_t$ of the $R$ values we have for point $t$ are a good representation of this data.

%%%fig 7

To evaluate the quality of our fit, we looking for parameter values which minimised the value of the {chi-squared function}
(for instance see  \cite{PTVF92})
\beq
 \chi^2 = \sum_{t,r} \frac{(y_{tr}-f_t)^2}{(\sigma_t)^2}
\eeq
%\cite[section 15.2, equation 15.2.2]{PTVF92}
where $y_{tr}$ is a path length of the node created at time $t$ as measured in run $r$. The fitting function used in the main text is $f_t = a \ln(t) + b$ but we also tried $f_t = a \ln(t) + b + (c/t)$ where $a,b,c,$ are constants to be found by fitting.

For numerical convenience (minimising memory requirements), for each of the path length we kept track of two values for each node occurring at $t$, the total sum of values $T_t$ and the sum of squares $S_t$, as given by
\beq
 T_t = \sum_r y_{tr},\quad S_t = \sum_r (y_{tr})^2
 \, .
\eeq

We do not expect the fit to be valid for small system sizes. For instance, in most of our work the initial graph is a complete DAG (see \appref{ainit} for a discussion of initial graph effects) in within this initial the path lengths scale as $y_t=\ellmax(t)=\ellgr(t)=t$. To deal with this we fit our data from time $t_0$, an additional parameter in our fit, to the largest time which is equal to the number of nodes $N$. So, in terms of the totals and the squares of totals, our chi-squared function becomes
\beq
 \chi^2
 =
 \sum_{t=t_0}^N
 \frac{ S_t -2T_t f_t+R(f_t)^2 }{ (S_t/R) - (T_t/R)^2}.
\eeq
Under our assumptions of normal fluctuations and independence as noted above, the probability of obtaining a particular $\chi^2$ value is given by the integral of the complementary cumulative distribution function of the $\chi^2$ distribution for the given degrees of freedom. The number of degrees of freedom in the model is equal to $(N-t_0+1).R$. We varied the $t_0$ parameter between $100$ and $10,000$ but found that this cutoff parameter $t_0$ had no significant influence on the resulting fits. This is to be expected, as even the largest $t_0$ value considered, $10,000$, constitutes a mere $0.1\%$ of the data we were using. So, in our work we used a fixed value of $t_0=1,000$.

The errors in the parameters of the linear fit to $f_t = a \ln(t) + b$ come from the square root of the diagonal elements of the covariance matrix. We found that there was no consistent pattern in the next-to-leading order terms, that is in $b$, so these are not shown. As noted elsewhere, this constant factor $b$ in our fit is likely to be effected by the form of the initial graph since this can add a significant constant to the lengths scales and the initial graph will effect lengths scales later in different ways depending on the parameters chosen. We also tried looking at higher order terms, using non-linear fits to  $f(t) = a \ln(t) + b + (c/t)$, but again saw no improvements in our results. For the results of the comparison  between the linear and non-linear fits see Table \ref{tlinnonlincompare}.

\begin{table*}[!htb]
\centering
\resizebox{\textwidth}{!}{%
\begin{tabular}{c|c|c|c|c|c}
m
  & $\alpha$
  & $\chi^2_{\mathrm{gr,lin}}-\chi^2_{\mathrm{gr,nonlin}}$
  & $\frac{\chi^2_{\mathrm{gr,lin}}-\chi^2_{\mathrm{gr,nonlin}}}{\chi^2_{\mathrm{gr,lin}}}\times 10^{7}$
  & $\chi^2_{\mathrm{max,lin}}-\chi^2_{\mathrm{max,nonlin}}$
  & $\frac{\chi^2_{\mathrm{max,lin}}-\chi^2_{\mathrm{max,nonlin}}}{\chi^2_{\mathrm{max,lin}}}\times 10^{7}$ \\
\hline\hline
2 & 1        & 1914.679   & $1.8958$ & 52115.2  & 51.6   \\
2 & 2        & 5500.539   & $5.4468$    & 47292.36     & 46.9    \\
2 & 3        & 10679.31      & $10.574$  & 32425.11  & 32.1 \\
2 & 4        & 15793.18  & $15.638$     & 11345.38         & 11.2  \\
3 & 1        & 4128.344   & $4.0881$   & 53148.67     & 52.7   \\
3 & 2        & 10891.1      & $10.784$    & 30573.03    & 30.3    \\
3 & 3        & 18021.77 & $17.849$        & 14582.06   & 14.5    \\
3 & 4        & 29150.27    & $28.865$  & 535.4066   & $0.531$    \\
4 & 1        & 6381.005   & $6.3185$    & 35694.73     & 35.4   \\
4 & 2        & 15188.28   & $15.041$ & 13707.66   & 13.6   \\
4 & 3        & 30406.43  & $30.109$     & 10.1752   & $0.0101$   \\
4 & 4        & 41108.91   & 40.72   & 3412.088    & $3.39$  \\
5 & 1        & 7572.032   & $7.4983$   & 31364.85   & 31.1     \\
5 & 2        & 23749.52    & 23.52  & 3088.395    & $3.06$   \\
5 & 3        & 36774.17  & 36.424   & 1128.841    & $1.12$       \\
5 & 4        & 66946.58  & 66.314 & 47772.77 & 47.3   \\
6 & 1        & 11489.57   & 11.378   & 22020.32       & 21.8     \\
6 & 2        & 26992.08   & 26.731 & 149.04        & $0.148$\\
6 & 3        & 61386.35      & 60.782        & 45868.17 & 45.4 \\
6 & 4        &   89429.041&  88.563       &        93100.79  &   92.2

\end{tabular}%
}
\caption{Absolute and relative difference between chi-squared values obtained by fitting path data to a linear function, (3.1), $\chi^2_{\mathrm{lin}}$ and to a non-linear function $f(t) = a \ln(t) + b + (c/t)$, $\chi^2_{\mathrm{nonlin}}$. The non-linear fit results in smaller $\chi^2$ values, but the improvement is marginal.  }
\label{tlinnonlincompare}
\end{table*}

%\tcomment{How we fit the data. Averaging over runs. Linear or non-linear fit. Low $t_0$ cutoff as displayed in  \cite{C19} Figs.\ 9,10,11,12. Method to find errors in fit parameters.}where $(a,\gamma,c,t_0)$ are the four parameters of our fit. An example of the fit and the data compared for one set of parameters is shown in \figref{foneparamfit}.

% ..................................................
\subsection{Initial graph effect}\label{ainit}

%In our simulations, we choose an initial graph arbitrarily. It seems that the effect of the type of initial graph we start the growth of the network with is not insignificant.

Throughout this work, we started our simulations from an initial graph which was the directed complete graph composed of $(2m+1)$ nodes. That is $\Gcal_\mathrm{compl}(t)$ where $\Vcal_\mathrm{compl}(t)=\{1, 2 \ldots, t\}$ and $\Ecal_\mathrm{compl}(t)= \{(s,r) | 1\leq s < r \leq t \}$. This is a dense, transitively complete graph but at each step in our simulation we always have exactly $m$ more edges than nodes which is an assumption in the analytical work.  A possible drawback is that this initial graph has long path lengths, $\ellgr(t)=\ellmax(t)=(t-1)$ for nodes in this graph. As both our longest and reverse greedy paths will start with paths in the initial graph, these long initial graph path lengths will produce a significant addition to those we measure. For instance, for $m=5$ this initial graph contributes up to 10 to any path we measure while the typical length scales we measure at late times are less than 100 in general.

We can, however, choose alternative initial graphs. To match the standard Price model and our analysis, we confine ourselves to initial graphs which are weakly connected and which have a single global source node\footnote{In fact, if we have several global source nodes we could add just one more ``global parent node'' with an edge from this extra node to all of the original global source nodes.  This adds one to all all the lengths we measure but creates a DAG with a single node of $\kin=0$.}
where $\kin=0$. A transitively reduced  \cite{CGLE14} version of our chosen initial graph is a single chain of nodes, so $\Gcal_\mathrm{chain}(t)$ where $\Vcal_\mathrm{chain}(t)=\{1, 2 \ldots, t\}$ and $\Ecal_\mathrm{chain}(t)= \{(r,r+1) | 1 \leq  r < t \}$, would also be a valid choice. This type of initial graph is as sparse as possible, all nodes have $\kin$ and $\kout$ equal to zero or one, but this initial graph has the same long path lengths as the complete graph. The reason the chain and the complete graph lead to different behaviour in the shift in the average path length at long times is because of their different initial degree distributions which alters the pattern of attachment at early times. That in turn alters the likelihood that the paths we measure join the initial graph at a particular node. In particular, we expect that the paths we measure in models starting with the complete graph are more likely to contain a smaller fraction of the initial graph as the earliest nodes will have higher out degrees but add shorter path lengths to our measurements. By way of contrast, in the chain, most nodes have the same initial degree, they are likely to have similar degree over time so the paths we measure are more likely to leave the initial graph at a later node so giving a longer contribution to the path length coming from this initial chain graph.

Alternatively, we could also use an initial graph, in which the single global source node points to the remaining nodes, which in turn are not pairwise connected. This is a ``directed star graph'' $\Gcal_\mathrm{star}(t)$ where $\Vcal_\mathrm{star}(t)=\{1, 2 \ldots, t\}$ and $\Ecal_\mathrm{star}(t)= \{(1,r) | 1 <  r \leq t \}$.
Now the degree distribution is highly skewed, with the central node  going to dominate attachment through the the cumulative advantage mechanism.  However, connecting to one of the other nodes only adds one to the paths we measure so we expect our models using this initial star graph will have the smallest paths lengths of the three cases considered.

\begin{figure*}[!ht]
    \centering
    \includegraphics[width = 1\linewidth]{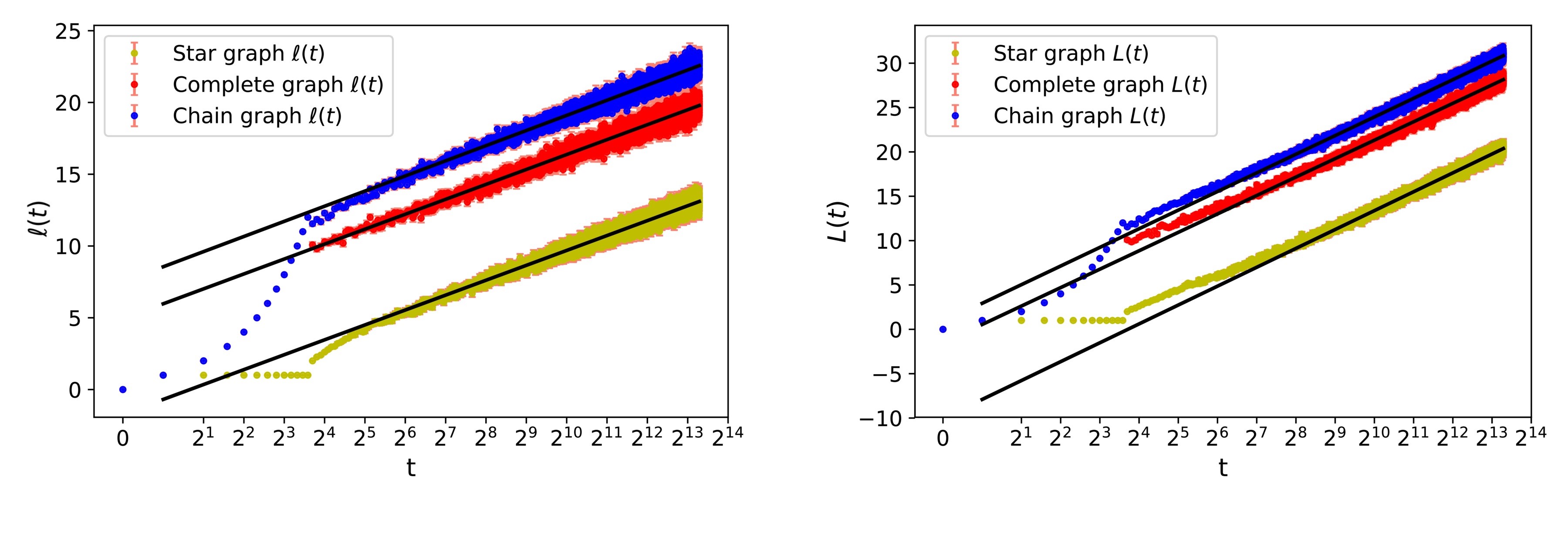}
    \caption[Path length scaling with t]
    {Path length (greedy on the left and the longest on the right) vs.\ $\ln(t)$ for 100 runs with m=6, fitness=2, $N=10^5$ using different initial graphs, all composed of $2m+1$ nodes.
    Fitted lines are of the form $a \ln(t) +b$.
    The fit parameter $a$ seems to be unaffected by the initial graph,
    whereas the fit parameter $b$ varies significantly, see Table \ref{t_initgraph_effect}.
    This causes the slope to remain stable but shifts the graphs vertically.
    The fit was obtained for nodes created from $t=1,000$ onwards.}
\label{finitgraph}
\end{figure*}

As \figref{finitgraph} shows, the results obtained using either of the three studied variants of the initial graph are distinguishable.
The asymptotic scaling of the path lengths as $a\ln{t}$ remains the same with the same value of the parameter $a$.  However the constant contribution, $b$ of (3.1), seems to depend on the choice of the initial graph.  As \figref{finitgraph} and \tabref{t_initgraph_effect} show, the $b$ coefficient behaves exactly as suggested above: the star graph gives the shortest path lengths while the chain gives the longest.

%%%fig 8

% BITMAP version

\begin{table}[hbt!]
\centering
\begin{tabular}{c|c|c|c}
%\multicolumn{3}{c }{Greedy path}\\
Path type&Initial graph& $a$&$b$\\
\hline\hline
Greedy&Star& $1.4973 \pm 0.0052$& $-0.688 \pm 0.04417$\\
Greedy&Complete& $1.5001 \pm 0.0038$& $5.9741\pm 0.0309$\\
Greedy&Chain& $1.5213 \pm 0.0035$ & $8.5548 \pm 0.0290$\\
Longest&Star& $3.0701 \pm 0.0056$ & $-7.9019 \pm 0.0476$\\
Longest&Complete& $2.9970 \pm 0.0034$ & $0.5458 \pm 0.0279$\\
Longest&Chain& $3.0303 \pm 0.0034$ & $2.9345 \pm 0.02789$
%\multicolumn{3}{c}{The longest path}\\
\end{tabular}
\caption{Variation in obtained fitting parameters for the greedy path scaling and the longest path scaling
when using various initial graphs in the Price model with $m=6$,
$\alpha=2$. The data from $t=1,000$ to $t=10^5$ was fitted to the
form $a \ln(t) +b$. The initial graph seems to have a small effect
on the slop, parameter $a$, but causes significant changes in the
intercept, parameter $b$.}
\label{t_initgraph_effect}
\end{table}

% ..................................................
\subsection{Next-to-leading Order Coefficient}

\begin{figure}[!htb]
    \centering
    \includegraphics[width=0.45\textwidth]{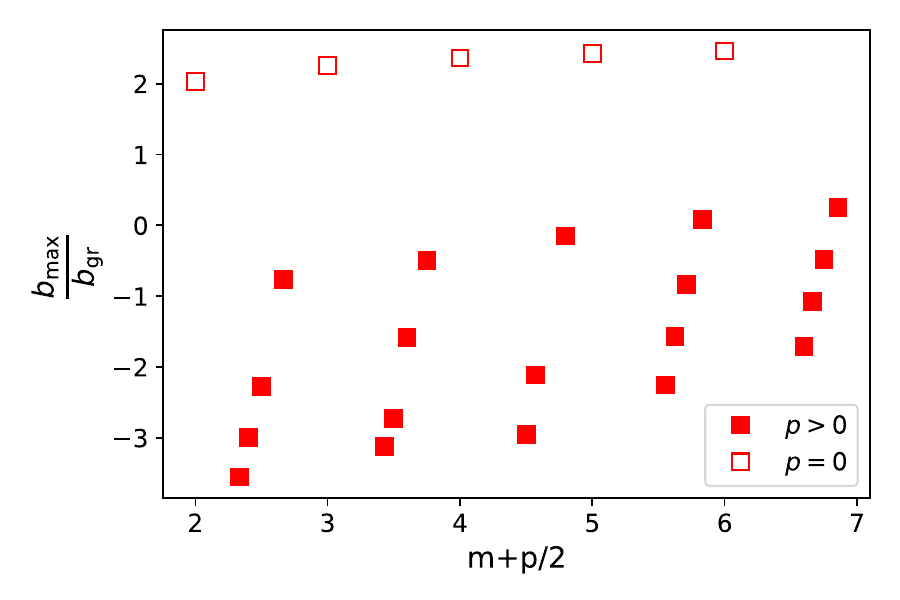}
    \caption{The ratio of $b_{\textrm{max}}$ and $b_{\textrm{gr}}$
    where $b$ is the next-to-leading order coefficient in the best fit
    of the numerical path length data to $a \ln(t) + b$:
    $b_{\textrm{max}}$ for the longest path data and $b_{\textrm{gr}}$ for the reverse greedy path data.
    These values were obtained by fitting the form to nodes created between $t=1,000$ and $t=10^8$ from 100 realisations.
    The errors on the fitted values of $b$,
    as estimated from the covariance matrix of the linear fitting algorithm,
    were smaller than the marker size so the uncertainties are not shown.}
    \label{fallparamfit_b}
\end{figure}

The next-to-leading order coefficient, $b$ of (3.1) did not reveal as clear trends as the coefficient $a$ of the same equation. There does not seem to be an obvious relation between $b_{\textrm{max}}$ and $b_{\textrm{gr}}$, as \figref{fallparamfit_b} shows.
%%%fig 9

% ..................................................
\subsection{Height Antichain Properties}\label{anumantichain}

\begin{figure*}[hbt!]
    \centering
    \includegraphics[width=\textwidth]{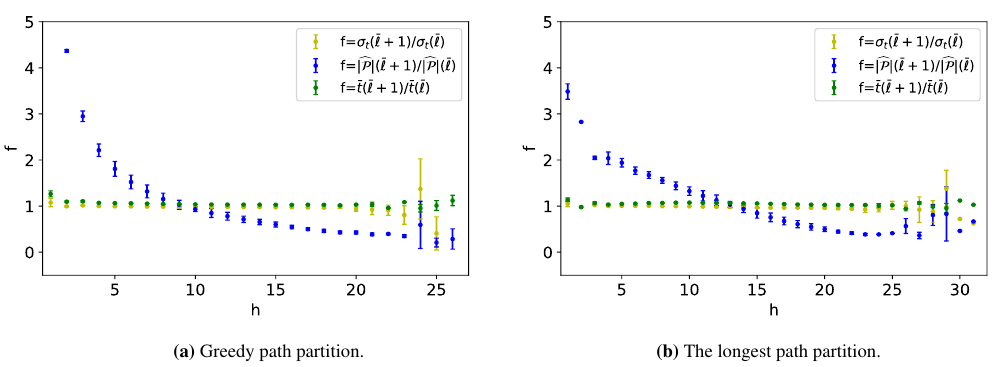}
%   % \begin{subfigure}[b]{0.475\textwidth}
%    %    \centering
%        \includegraphics[width=0.45\textwidth]{partition_scaling_m_eq_2_a_eq_1_no_networks_10_networksize_1000000_greedy}
%  %      \caption[Network2]%
%  %      {{\small Reverse greedy path partition.\\}}
%  %      \label{sgrpart}
%  %  \end{subfigure}
%  %  \hfill
%   % \begin{subfigure}[b]{0.475\textwidth}
%%        \centering
%        \includegraphics[width=0.45\textwidth]{partition_scaling_m_eq_2_a_eq_1_no_networks_10_networksize_1000000_longest.png}
% %       \caption[]%
% %       {{\small Longest path partition.\\}}
%  %      \label{smaxpart}
%  %  \end{subfigure}
        \caption[Path length scaling with t]
    {Ratio of quantities related to vertex partitions: the average time of nodes in the partition $\tbar(\ellgrbar)$, the standard deviation of times $\sigma_t(\ellgrbar)$ and the number of nodes in the block for adjacent blocks $ |\Pcalhat|(\ellgrbar)$ for greedy path partition on the left and for the height antichains on the right.} %Left figure shows results for the greedy path partition, right figure shows results for the longest path partition.}
\label{fpartitionscaling}
\end{figure*}

The \tdef{height} of a node in a DAG is the length of the longest path to a node from any global source node, any node with zero in-degree.
Thus in the Price model, the height of a node is simply the longest path length from the initial node to any node, our $\ellmax$.
We can also define \tdef{antichains}, sets of nodes which are not connected by any path, for example see  \cite{VE20}.
A simple example in a DAG is a \tdef{height antichains}, a set of nodes which are all at the same height since nodes connected by a path cannot be of the same height.
Based on the $\ln(t)$ scaling of the reverse greedy path length, we can speculate that the average height $\ellmax$ of a node $t$ will scale as  $t = \tilde{t}_\labelmax (\mu)^\ellmax $.
Then, as discussed in \secref{avpart}, we find that various properties of the antichains should scale in a simple way with height:
the average time of nodes in the partition $\tbar(\ellmax)$  as in \eqref{atbargr},
the standard deviation of the times of nodes in the partition $\sigma_t(\ellmax)$ as in \eqref{asigmagr},
and the number of nodes in the block $|\Pcalhat|(\ellmax)$   as in \eqref{apartpropgr}. Results for ratio of these quantities for adjacent blocks are shown in \figref{fpartitionscaling}.% \figref{afmeasureheightac} and \figref{afheightacratios}.

\clearpage

\end{document}